\documentclass[12pt,preprint]{aastex}

\def\lcdm{$\Lambda$CDM}
\def\om{\Omega_m}
\def\oma{\omega_{0.3}}
\def\s8{\sigma_8}
\def\ml{$M/L$}
\def\mlmean{$\langle M/L\rangle$}

\def\mlclnormp{$(M/L)_{\rm cl} \div \langle M/L\rangle$}
\def\mlcl{$(M/L)_{\rm cl}$}
\def\mlclp{$(M/L)_{\rm cl}$}
\def\mla{$M/L_{18}$}
\def\mlb{$M/L_{20}$}
\def\mlB{$(M/L)_{B}$}
\def\mlcla{$(M/L_{18})_{\rm cl}$}
\def\mlclb{$(M/L_{20})_{\rm cl}$}
\def\etal{et\,\,al.}
\def\hmpc{$h^{-1}\,$Mpc}
\def\hkpc{$h^{-1}\,$kpc}
\def\xg{$\xi_g(r)$}
\def\clf{$\Phi(L|M_h)$}
\def\x2{$\chi^2$}
\def\hmsol{$h^{-1}\,$M$_\odot$}
\def\whmsol{$\omega_{0.3}\,h^{-1}\,$M$_\odot$}
\def\wp{$w_p(r_p)$}
\def\gad{{\scriptsize GADGET}}

\def\hml{$h\,$M$_\odot/$L$_\odot$}

\def\whml{$\omega_{0.3}\,h\,$M$_\odot/$L$_\odot$}

\def\kd{$k_\delta$}
\def\kms{km\,s$^{-1}$}

\def\plin{P_{\rm lin}(k)}

\def\mmin{M_{\rm min}}
\def\mlim{M_{\rm lim}}
\def\mlima{M_{\rm lim,1}}
\def\mlimb{M_{\rm lim,2}}
\def\mcut{M_{\rm cut}}
\def\navg{\langle N\rangle_M}
\def\navgmh{\langle N\rangle_{M_h}}
\def\navga{\langle N\rangle_{M_1}}
\def\navgb{\langle N\rangle_{M_2}}
\def\nsat{\langle N_{\mbox{\scriptsize sat}}\rangle_M}

\def\ncen{\langle N_{\mbox{\scriptsize cen}}\rangle_M}
\def\ngal{N_{\mbox{\scriptsize gal}}}
\def\nbcg{N_{\mbox{\scriptsize BCG}}}
\def\rmax{r_{\mbox{\scriptsize max}}}
\def\plum{\rho_{\mbox{\scriptsize lum}}}
\def\pcrit{\rho_{\mbox{\scriptsize crit}}}

\def\ngavg{\bar{n}_g}
\def\ngavgp{\bar{n}^\prime_g}

\def\NNm1{\langle N(N-1) \rangle}
\def\xis{\xi_{\rm 1h}}
\def\xid{\xi_{\rm 2h}}
\def\Rvir{R_{\rm vir}}
\def\intdn{\int_{0}^\infty dM\frac{dn}{dM}} 
\def\intdnM{\int_{0}^{M_{\rm lim}} dM\frac{dn}{dM}}

\begin{document}

\title{On The Mass-to-Light Ratio of Large Scale Structure } 
\author{Jeremy L. Tinker\altaffilmark{1}, 
David H. Weinberg\altaffilmark{1}, 
Zheng Zheng\altaffilmark{2}\altaffilmark{,4}, \&
Idit Zehavi\altaffilmark{3}}

\altaffiltext{1}{Department of Astronomy, The Ohio State University,
140 W. 18th Avenue, Columbus, Ohio 43210}

\altaffiltext{2}{Institute for Advanced Study, School of Natural
Sciences, Einstein Drive, Princeton, NJ 08540}

\altaffiltext{3}{Steward Observatory, University of Arizona, 933 N. Cherry
Avenue, Tucson, AZ 85721}

\altaffiltext{4}{Hubble Fellow}

\begin{abstract}

We examine the dependence of the mass-to-light (\ml) ratio of
large-scale structure on cosmological parameters, in models that are
constrained to match observations of the projected galaxy correlation
function \wp\ and the galaxy luminosity function. For a sequence of
cosmological models with a fixed, observationally motivated power
spectrum shape and increasing normalization $\s8$, we find parameters of
the galaxy halo occupation distribution (HOD) that reproduce \wp\
measurements as a function of luminosity from the Sloan Digital Sky
Survey (SDSS). From these HOD models we calculate the $r$-band
conditional luminosity function \clf, and from this the mean
\ml\ ratio as a function of halo mass $M_h$. We also use \clf\ to
populate halos of N-body simulations with galaxies and thereby compute
\ml\ in a range of large-scale environments, including cluster infall
regions. For all cosmological models, the \ml\ ratio in high mass
halos or high density regions is approximately independent of halo
mass or smoothing scale.  However, the ``plateau'' value of \ml\
depends on $\s8$ in addition to the obvious proportionality with the
matter density parameter $\om$, and it represents the universal value
\mlmean $\,= \om \pcrit/\plum$ only for models in which the galaxy
correlation function is approximately unbiased, i.e., with $\s8
\approx \s8{_g}$.  Our results for cluster mass halos follow the trend
\mlcl $\,= 577\,(\om/0.3)\,(\s8/0.9)^{1.7}$ \hml. Combined with
Carlberg et al.'s (\citeyear{carlberg96}) mean \ml\ ratio for CNOC
galaxy clusters, this relation implies
$(\s8/0.9)\,(\om/0.3)^{0.6}=0.75\pm 0.06$.  \ml\ estimates for SDSS
clusters and the virial regions of clusters in the CAIRNS survey imply
a similar value of $\s8\om^{0.6}$, while the CAIRNS estimates for
cluster infall regions imply a lower value.  These results are
inconsistent with parameter values $\om\approx 0.3$, $\s8\approx 0.9$
favored by recent joint analyses of cosmic microwave background
measurements and other large scale structure data, though they agree
with values inferred from van den Bosch et al.'s (\citeyear{vdb03})
analysis of the 2dF Galaxy Redshift Survey. We discuss possible
resolutions of this discrepancy, none of which seems entirely
satisfactory.  In appendices we present an improved formula for halo
bias factors calibrated on our $360^3$ N-body simulations and an
improved analytic technique for calculating the galaxy correlation
function from a given cosmological model and HOD.

\end{abstract}



\section{Introduction}

Determining the matter density of the universe is one of the key goals
of observational cosmology.  By definition, the mean matter density is
the product of the mean luminosity density and the mean mass-to-light
ratio of the universe, \mlmean, making the density parameter

\begin{equation}
\label{e.ml_omega}
\Omega_m = \langle M/L\rangle \times \plum/\pcrit.
\end{equation}

\noindent One of the classic methods of inferring $\om$ is to estimate
the \ml\ ratios of galaxy groups or clusters, then multiply by the
mean luminosity density derived from the galaxy luminosity function
(e.g., \citealt{gott74}; \citealt{peebles86}; \citealt{bahcall95};
\citealt{carlberg96}). However, dynamical methods of estimating \ml\
necessarily focus on dense regions of the galaxy distribution, so this
route to $\om$ relies on the assumption that the galaxy population in
these region is representative of the universe as a whole. Observed \ml\
ratios rise steadily from the scale of binary galaxies to groups to rich
clusters, but there is an approximate plateau in \ml\ values for the
richest bound systems, and the mass-to-light ratios inferred from the
dynamics of superclusters and the extended infall regions around
clusters extend this plateau to larger scales (see, e.g.,
\citealt{bahcall95}; \citealt{bahcall00}, hereafter B00;
\citealt{rines04}). The existence of this plateau is sometimes taken as
evidence that the measured values of \ml\ do indeed represent the
universal value.

This paper has two goals. The first is to assess the above line of
reasoning: does the existence of a plateau in \ml\ at large scales
imply that the universal value has been reached? The second is to
assess the implications of observed \ml\ ratios for the values of
$\om$ and the matter fluctuation amplitude $\s8$, which we will show
plays a crucial role. (Here $\s8$ is the rms matter fluctuation in
spheres of radius 8 \hmpc, calculated from the linear matter power
spectrum, where $h\equiv H_0/100$ \kms\,Mpc$^{-1}$.) The key to these
assessments is a model for the relation between galaxies and dark
matter (a.k.a. galaxy bias) that extends to the non-linear regime. We
will derive this relation empirically, by fitting models of the halo
occupation distribution (HOD; see \citealt{bw02} and references
therein) to the projected correlation functions of galaxies in the
Sloan Digital Sky Survey (SDSS; \citealt{york00}), along the lines of
Zehavi et al. (2004a,b). HOD models describe bias at the level of
virialized halos by the probability $P(N|M)$ that a halo of virial
mass $M$ contains $N$ galaxies of a specified type, together with
prescriptions for spatial and velocity biases within halos.

B00 also addressed the complications of mass-to-light ratios, drawing
on the results of a hydrodynamic simulation of the galaxy
population. Their simulation predicts a plateau in the $B$-band \ml\
that is close to the universal value, except for a modest
``anti-bias'' arising from the older stellar populations of galaxies
in dense environments. Applying this anti-bias correction to the
observed \mlB\ values, they infer $\om=0.16\pm 0.03$. However, B00
considered only a single cosmological model, with $\s8=0.8$. We will
show that, when the bias relation is inferred empirically by matching
the observed correlation function, a plateau in \ml\ at high halo
masses or overdensities is a generic result. However, the plateau
corresponds to the true universal \ml\ {\it only} for models with
$\s8\approx \s8{_g}$, where $\s8{_g}$ is the rms galaxy count fluctuation. For lower
$\s8$, galaxies must be positively biased to match the observed
clustering, and the \ml\ plateau is below the universal value. For
higher $\s8$, the galaxy population is anti-biased in the dense
regions, and the \ml\ plateau is higher than the universal value.

\cite{zehavi04b} infer HOD parameters as a function of galaxy $r$-band
luminosity by fitting measurements of the projected galaxy correlation
function \wp, assuming a cosmological model with $\s8=0.9$. We carry out
a similar procedure, but we consider a range of $\s8$ values from 0.6 to
0.95. We keep the shape of the linear matter power spectrum $\plin$
fixed, regardless of $\s8$ or $\om$, because it is well constrained
empirically by the combination of microwave background anisotropies and
large-scale galaxy clustering measurements (e.g., \citealt{percival02};
\citealt{spergel03}; \citealt{tegmark04}), and because the physical
effects of changes in $\om$ are much more transparent if they are kept
separate from $\plin$ changes\footnote{ In the inflationary cold dark
matter (CDM) framework, a change in $\om$ with all other parameters held
constant alters the shape of $\plin$. However, variations other
parameters, such as the Hubble constant, the inflationary spectral
index, spectral running, the baryon density, and the neutrino mass can
compensate for those changes, at least to some degree.}. For a specified
$\om$ and $\s8$, we can calculate \ml\ as a function of halo mass
analytically, given the HOD parameters derived by fitting
\wp\ (see \S 2). To compute \ml\ over larger scales, we use the
derived HODs to populate the halos of N-body simulations.

As shown by \cite{ztwb02}, for a fixed linear power spectrum, a change
to $\om$ simply shifts the mass and velocity scales of virialized
halos, with negligible effect on the shape of the halo mass function
or the clustering of halos at a given (scaled) mass. For two
cosmological models that differ only in the value of $\om$, we can
obtain identical real-space galaxy clustering by simply shifting the
HOD mass scale in proportion to $\om$. With bias constrained by the
observed clustering, therefore, our predicted \ml\ ratios for a given
$\s8$ are simply proportional to $\om$. We will often indicate this
scaling by quoting predicted \ml\ ratios in units of \whml, where
$\oma=\om/0.3$. The scaling with $h$ arises in the observations
because inferred dynamical masses scale as $h^{-1}$ and luminosities
as $h^{-2}$. The same $h$ scaling arises in the predictions because
the characteristic mass in the halo mass function is proportional to
$h^{-1}$ at fixed $\om$ and $\s8$ and we automatically impose the
observed luminosity scale, which is proportional to
$h^{-2}$. Uncertainties in the value of $h$ therefore have no impact
on our conclusions.

Most recent applications of the mass-to-light method yield $\om\approx
0.15-0.20$ (e.g., \citealt{bahcall95}; \citealt{carlberg96}; B00;
\citealt{lin03}; \citealt{rines04}). Alternative methods of inferring
$\om$, from cosmic microwave background (CMB) anisotropies, the shape
of the galaxy power spectrum, Type Ia supernovae, and the
baryon-to-total mass ratio in clusters, have begun to converge on a
significantly higher value of $\om \approx 0.3$ (e.g. Turner
2002). Recently Tegmark \etal\ (2004) combined WMAP CMB data with the
SDSS galaxy power spectrum to infer $\om=0.30 \pm
0.04$. \cite{seljak04} combined these data with the SDSS
Lyman-$\alpha$ forest spectrum to obtain $\om = 0.28\pm0.02$. The
tension between \ml\ estimates of $\om$ and the higher values from
other methods warrants a careful investigation of the assumptions that
underly the \ml\ approach. We will show that the conflict can be
alleviated if $\s8$ is low, $\sim0.6-0.7$. However, the discrepancy
persists for a matter fluctuation amplitude of $\s8\approx 0.9$, which
is favored by the \cite{tegmark04} and \cite{seljak04} analyses.

Our approach is similar in spirit to the conditional luminosity function
(CLF) analyses of clustering in the Two-Degree Field Galaxy Redshift
Survey (2dFGRS; \citealt{colless01}) by Yang et al. (2003; 2004),
\cite{vdb03}, and \cite{mo04}. In practice there are important
differences in the two techniques. In the CLF, the luminosity function
within an isolated halo is parameterized by a Schechter function, while
in our approach this function is determined non-parametrically. We fit
the full projected correlation function, while the above CLF papers fit
the luminosity dependence of the galaxy correlation length, together
with the galaxy luminosity function. Despite the many differences in
analysis and parameterization, and the use of independent galaxy
clustering measurements derived from an $r$-band selected sample instead
of a $b_J$-selected sample, we reach a bottom line similar to that of
\cite{vdb03}: obtaining cluster mass-to-light ratios close to the
observational estimates of $\sim 350$ \hml\
(e.g. \citealt{carlberg96}; B00) requires either $\om$ or $\s8$ to be
significantly below the currently favored values of 0.3 and 0.9 (see
\S 4).


\section{HOD Modeling of \wp}

The key ingredient in the HOD prescription of galaxy bias is $P(N|M)$,
the probability that a halo of mass $M$ contains $N$ galaxies of a
specified class. The galaxy classes we will consider are defined by
thresholds in $r$-band luminosity. Our model for $P(N|M)$ is motivated
by theoretical studies of the HOD using semi-analytic galaxy formation
methods, high-resolution N-body simulations, and full hydrodynamic
cosmological simulations (specifically \citealt{kravtsov04} and
\citealt{zheng04}; for earlier work see \citealt{jingmoborner98,kauffmann97}, 1999;
\citealt{benson00}; \citealt{seljak00}; \citealt{white01};
\citealt{yoshikawa01}; \citealt{berlind03}). This model distinguishes
halo central galaxies from satellites in halos containing multiple
galaxies. In our standard parameterization, the number of central galaxies above
the luminosity threshold changes sharply from zero to one at a minimum
halo mass $\mmin$. We also consider a model in which the probability
of hosting a central galaxy is $\exp(-\mmin/M)$, to test the
sensitivity of our results to the assumed sharpness of the central
galaxy threshold.

We adopt a functional form

\begin{equation}
\label{e.nsat}
\nsat = \exp \left(-\frac{\mcut}{M-\mmin}\right) \left( \frac{M}{M_1}\right)
\end{equation}

\noindent for the mean number of satellites in a halo of mass $M\ge
\mmin$, and $\nsat=0$ for $M<\mmin$. Here $\mcut$ is a cutoff
mass scale for the satellite galaxy power law, which allows a soft
transition to halo masses that host no satellites. \cite{zehavi04b}
showed that a model of this form allows good fits to the observed \wp\
for the concordance cosmology, though they mainly focused on an
alternative parameterization with a sharp cutoff in $\nsat$ and a free
power-law index $\nsat \propto M^\alpha$. The fixed-$\alpha$ with
varying $\mcut$ parameterization is better suited to our purposes here
because we will difference occupation functions for adjacent
luminosity thresholds to obtain $P(N|M)$ for luminosity bins, and
small statistical errors in $\alpha$ can drastically affect the number
of of satellites in individual bins at high halo mass.  We assume a
Poisson distribution of satellite number relative to the mean $\nsat$,
as implied by the theoretical studies.

This parameterization of $P(N|M)$, while restricted, has the virtue of
capturing theoretical predictions quite well while introducing only
three free parameters. An essentially perfect fit to theoretical
results requires two additional parameters providing freedom in
$\alpha$ and the central galaxy cutoff shape
(\citealt{zheng04}). However, we expect the three parameter model to
be adequate to our purposes here, predicting the mass-to-light ratios
in and around high mass halos.

We will investigate the effects of changing the assumed value of
$\alpha$, but we note that the choice of $\alpha=1$ is well motivated by
the results of semi-analytic models, hydrodynamic simulations, and
collisionless N-body simulations (\citealt{zheng04,
kravtsov04}). Observational estimates of $\alpha$ are subject to the
effects of interlopers, completeness limits, and uncertainty in the halo
mass estimates, and the agreement among observational estimates is not
as good as for the theoretical studies. \cite{lin04}, using a sample of
rich clusters from the 2MASS survey, find $N_{\rm sat} \propto
M^{0.87}$, in agreement with the much smaller sample of \cite{rines04},
who find $\alpha=0.84$. \cite{kochanek03}, who analyze a different 2MASS
cluster sample, find $\alpha=1.1$. Analysis of the 2dF
Percolated-Inferred Galaxy Group catalog (\citealt{eke04}) by
\cite{collister04}, which constitutes the largest group sample analyzed
to date, yields $\alpha=0.99$, a value consistent with the preliminary
results of the group multiplicity function for SDSS galaxies
(A. Berlind, private communication).

We fit the observed \wp\ for each luminosity threshold sample of SDSS
galaxies measured in \cite{zehavi04b}, as listed in their Table 2. For
the $M_r=-20$ threshold\footnote{Throughout this paper, we quote
absolute magnitudes for $h=1.0$.}, we use the sample limited to redshift
$z\le 0.06$. We fit each sample for five values of $\s8$: 0.95, 0.9,
0.8, 0.7 and 0.6. Our results for $\s8=0.9$ are similar to those of
\cite{zehavi04b}, except for the slight differences in
parameterization and calculational method (see Appendix B). For all
calculations, we adopt a linear theory matter power spectrum with
inflationary spectral index $n_{s}=1$ and shape parameter $\Gamma=0.2$
in the parameterization of \cite{ebw}. Satellite galaxies are
assumed to follow the ``universal'' halo density profile of Navarro,
Frenk, \& White (1997) with no internal spatial bias. The halo
concentrations are calculated in the same manner as Kuhlen \etal\
(2004) assuming $\om=0.3$, with the halo edge defined as the radius at
which the mean interior density is 200 times the mean value. The
calculation of \wp\ for a specified cosmology and HOD is described in
Appendix B. Our method is similar to that of \cite{zheng04a}, which
was used by \cite{zehavi04b}, but we implement an improved treatment
of halo exclusion and adopt more accurate halo bias factors
inferred from our N-body simulations (Appendix A).

The data for each luminosity threshold are fit by minimizing \x2\
using the full covariance matrix for each sample. The jackknife
estimates of these matrices are discussed in detail by
\cite{zehavi04b}. The number of free parameters in the fit is reduced
from three to two by matching the space density of galaxies for each
sample. In practice, this constraint is used to fix the value of
$\mmin$ for a given $\mcut$ and $M_1$. We ignore the uncertainty in
the space density itself. We require that the number density of
satellite galaxies between two adjacent magnitude limits be smaller
for the brighter sample at all halo masses, a consistency condition
that only affected the result in two of the 45 fits performed. The
results of the fits are listed in Table 1.

\begin{figure}
\epsscale{1.0}
\plotone{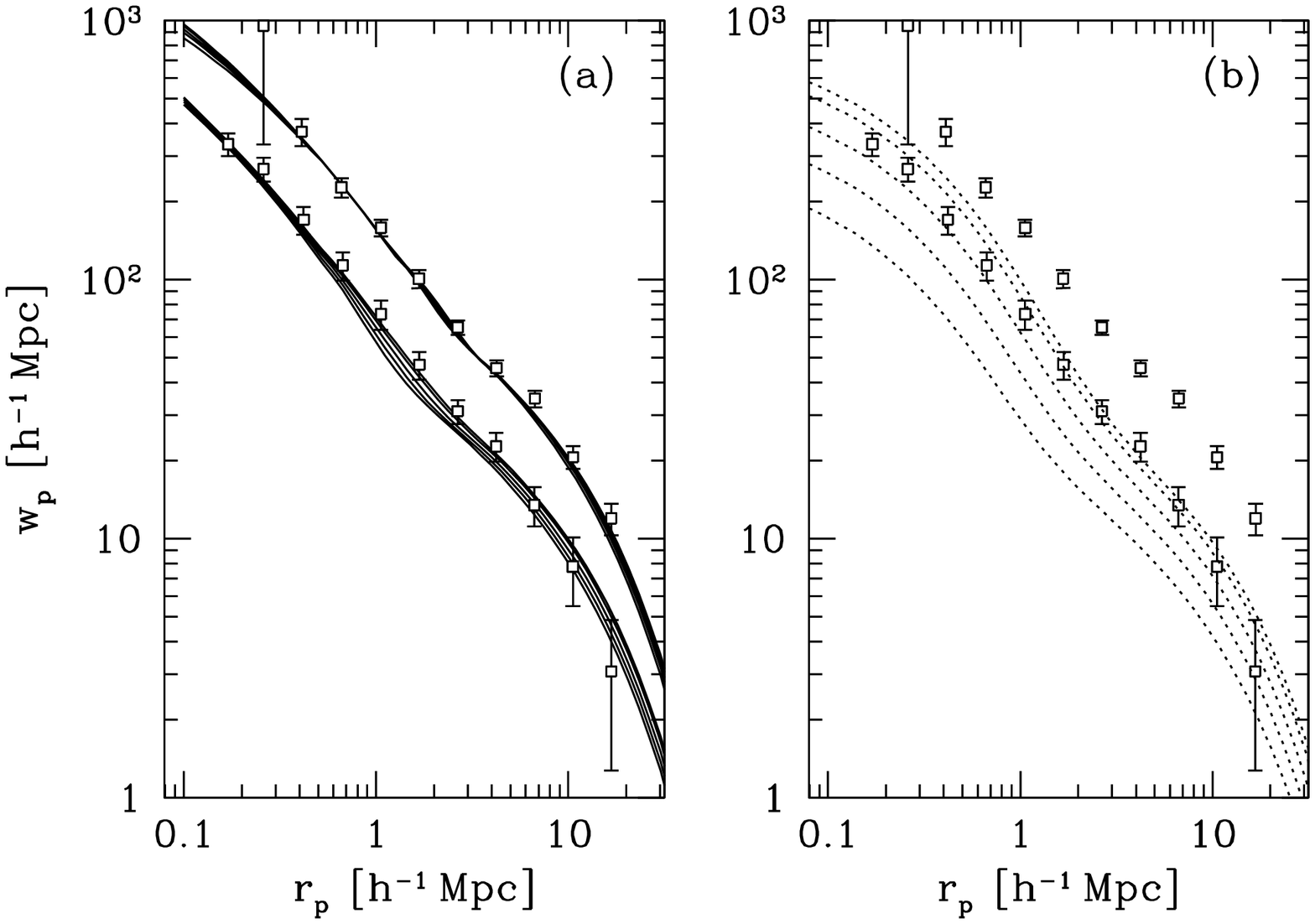}
\caption{ \label{wp_fits} (a) The HOD fits are compared to the \wp\
  data for galaxy samples $M_r<-20$ (lower points) and $M_r<-21.5$
  (upper points). The five solid curves represent the five values of $\s8$,
  with the lowest curve being $\s8=0.6$ in both fits, and each
  subsequent curve going in order of increasing $\s8$. (b) The
  projected dark matter correlation functions for $\om=0.3$ and all
  five values of $\s8$. The curves, from lowest to highest, go in
  order of increasing $\s8$. For comparison, the \wp\ data are plotted
  as well.}
\end{figure}

Figure \ref{wp_fits}a shows two examples of these fits. The observed
\wp\ for $M_r<-20$ and $M_r<-21.5$ are shown by the symbols with error
bars, plotted on top of the HOD fits for each value of $\s8$. For
$M_r<-21.5$, the different fits are difficult to distinguish. For
$M_r<-20$, the larger error bars allow the best-fit HOD models to
differentiate some at large scales, even though the $\chi^2$ values
per degree of freedom are all less than one. Note that strong
covariances between data points make a simple visual estimate of
$\chi^2$ unreliable. In Figure \ref{wp_fits}b, the data are plotted
against the nonlinear matter correlation functions for each value of
$\s8$. These correlation functions are calculated by Fourier
transformation of the \cite{smith03} non-linear $P(k)$, which takes
the linear power spectrum as input, and assume $\om=0.3$. The
differences in the matter distributions are large, but are easily
overcome even with our restrictive three-parameter HOD.

The values of $\chi^2$ per degree of freedom for the HOD fits
fluctuate from sample to sample, with a median value below 1.0 for
$\s8\ge 0.8$ and above 1.0 for $\s8\le 0.7$. We do not regard the
somewhat higher $\chi^2$ values as significant arguments against low
$\s8$, however, because our three-parameter description of the HOD is
quite restrictive. In addition, adopting a CMBFAST (\citealt{cmbfast})
power spectrum with the \cite{tegmark04} cosmological parameters in
place of the $\Gamma=0.2$, \cite{ebw} power spectrum has a
noticeable effect on $\chi^2$ values, though it makes almost no
difference to the best-fit HOD parameters themselves. Adding freedom
to the HOD parameterization or changing the input power spectrum could
thus have significant effect on the quality of the \wp\ fits, but we
expect only minor influence on our mass-to-light ratio predictions
because the overall shape of $\navg$ is well constrained by the number
density and \wp\ measurements.

From the HOD fits listed in Table 1, we can calculate a discrete
estimate of the conditional luminosity function, \clf, the luminosity
function of galaxies in halos of mass $M_h$. Since the HODs describe
samples brighter than a specified absolute magnitude, differencing
the number of galaxies in a halo of mass $M_h$ within the magnitude
bin of width $\Delta M_r=0.5$ mag yields \clf\ within that bin:

\begin{equation}
\label{e.clf}
\Phi(L|M_h)\Delta L = \navgmh^{(M_r)} - \navgmh^{(M_r-\Delta M_r)}.
\end{equation}

\noindent We use luminosity rather than magnitude on the left hand
side of equation (\ref{e.clf}) for clarity of notation. In this form,
equation (\ref{e.clf}) is normalized such that a summation of \clf\
over all magnitude bins returns $\navg$ for the $M_r<-18$ sample. The
parameters in Table 1 assume our standard HOD model with a sharp
threshold for the central galaxy occupation. When we use an
exponential cutoff for the central occupation, we keep the same values
of $M_1$ and $\mcut$ but adjust $\mmin$ slightly so that the mean
galaxy space density remains fixed.

\begin{figure}
\epsscale{1.0}
\plotone{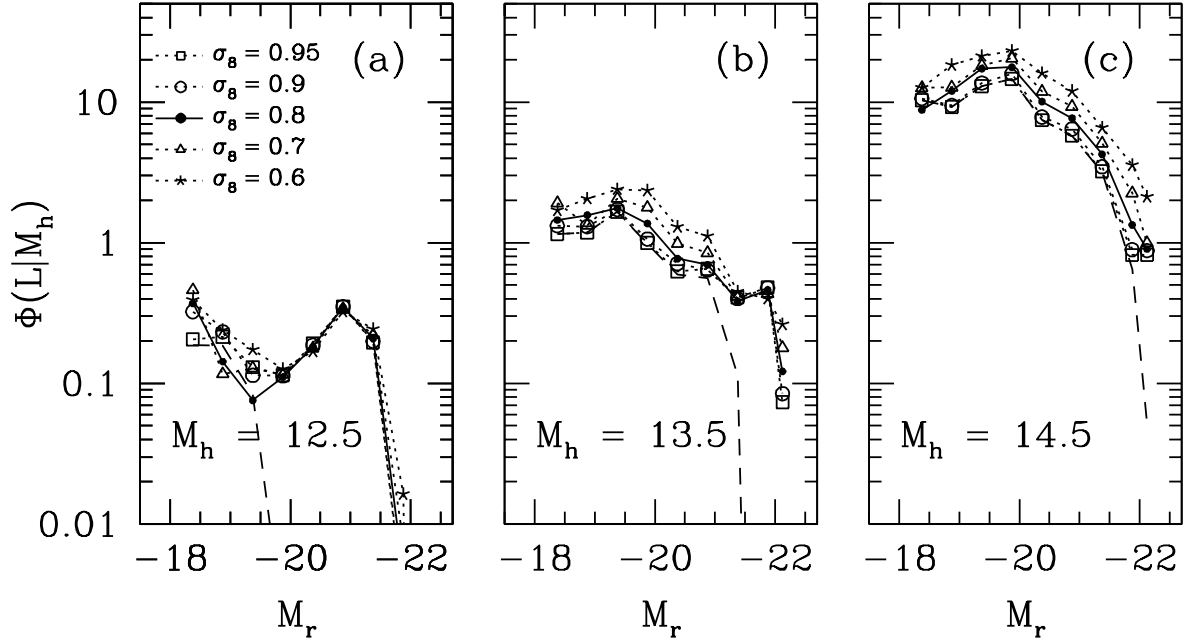}
\caption{ \label{clf} The conditional luminosity function, \clf, is
  plotted for three different halo masses, $\log\,(M\,h/{\rm M}_\odot) =
  12.5$, $13.5$,a dn $14.5$, and for all five values of $\s8$. The
  curves are all normalized such that the area under each curve is
  $\navg$ for $M_r<-18$. The dashed line in each plot is an example of
  the contribution to \clf\ by satellite galaxies only (here for
  $\s8=0.95$). At $M=3\times 10^{12}$ \hmsol, satellites contribute
  only to the faint end of \clf, while for the cluster-mass halos of
  panel (c), satellites dominate the luminosity function for all but the
  brightest magnitude bin. }
\end{figure}

Figure \ref{clf} plots the conditional luminosity functions for three
halo masses for the exponential central cutoff. Panel (a) shows \clf\
for $M_h = 3\times 10^{12}$ \hmsol, normalized so that the area under
each curve is the total number of galaxies expected at this halo
mass. The dashed line that falls rapidly at $M_r < -19.5$ is the
satellite galaxy contribution to \clf\ for the model with
$\s8=0.95$. At this low halo mass, the number of satellite galaxies
brighter than $M_r=-18$ is smaller than the number of central
galaxies, but satellites dominate at the faintest magnitudes. The
variation of \clf\ with $\s8$ is minimal because at this point on the
halo mass function the space density of halos is only weakly sensitive
to $\s8$. The difference in the total number of galaxies brighter than
$M_r=-18$ between the low and high values of $\s8$ is
$1.84-1.56=0.28$.

Panel (b) shows \clf\ for $M_h = 3\times 10^{13}$ \hmsol. Halos of
this mass are intermediate between those that host galaxy groups and
clusters, with $\navg\approx 8$ for $M_r\le -18$. Satellite galaxies
dominate the conditional luminosity functions out to $M_r = -21$,
beyond which the brighter central galaxies dominate. The different
values of $\s8$ now produce small differences in the overall
normalization of \clf. As $\s8$ decreases, the space density and
clustering amplitude of high mass halos decreases, and the mean
occupation of these halos must grow to keep the galaxy number density
and clustering amplitude constant. In panel (c), which plots \clf\ for
cluster-sized halos of $M_h = 3\times 10^{14}$ \hmsol, the differences
in the models are clear. The low-$\s8$ model has nearly twice as many
galaxies per halo as $\s8=0.95$. For all the models, \clf\ is
dominated by satellite galaxies at all but the brightest magnitude bin.

Our eventual qualitative conclusion is already evident from Figure
\ref{clf}. With the galaxy space density and clustering amplitude
fixed to match observations, a low-$\s8$ model must have a larger
fraction of its galaxies in massive halos, and it therefore predicts
lower mass-to-light ratios in these halos.

To quantify this point, we calculate mass-to-light ratios as a
function of halo mass by summing over luminosity,

\begin{equation}
\label{e.ml_halo}
M/L = M_h \left[ \sum_i L^{(i+1/2)}\,\left(\navg^{(i)} - \navg^{(i+1)}\right) \right]^{-1},
\end{equation}

\noindent where $i$ denotes the magnitude threshold, which runs from
$-18$ to $-22$. All galaxies in each magnitude bin are assumed to be
at the midpoint of the bin, and galaxies in the $M_r<-22$ sample are
assumed to have $M_r=-22.25$, an assumption that has negligible
influence on the results. Equation (\ref{e.ml_halo}) is equivalent to
$M/L = M_h/\int_{L_{\rm min}}^{L_{\rm max}} L\,\Phi(L|M_h)\,dL$ given
our discrete estimate of \clf\ in equation (\ref{e.clf}).

\begin{figure}
\epsscale{0.5}
\plotone{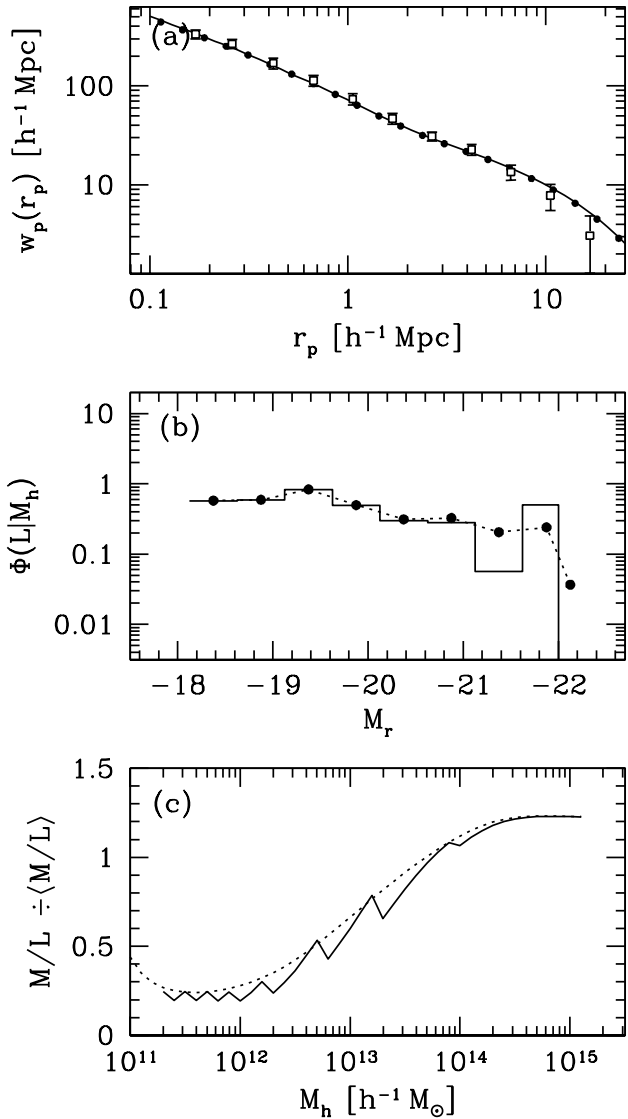}
\caption{ \label{hod_compare} A comparison between soft and hard
  central galaxy cutoffs. Panel (a) demonstrates that the different
  central occupation functions do not change the predicted \wp. The
  open squares are the data for $M_r<-20$ galaxies from
  \cite{zehavi04b}. The solid line is the HOD fit to the data using a
  hard cutoff. The solid circles represent the same HOD parameters, but \wp\
  is recalculated for a soft cutoff. Panel (b) shows the difference in
  \clf\ from the two methods for $M=3\times 10^{13}$ \hmsol. A hard
  cutoff, shown with the solid histogram, produces a significant bump at
  $M_r \sim -22$, which is smoothed out by the soft cutoff, shown by
  the dotted line and filled circles. Panel (c) shows the \ml\ ratio
  as a function of halo mass for the two methods. The solid line
  represents the hard cutoff and the dotted lines represents the soft
  cutoff. The differences are small, and nearly negligible at high
  masses. }
\end{figure}

Before proceeding to our main results, we want to check the
sensitivity of our calculations to the form of the central galaxy
cutoff and the value of $\alpha$. In Figure \ref{hod_compare}a,
points with error bars are the SDSS data for $M_r<-20$ from
\cite{zehavi04b}, and the solid line shows the fit with a hard cutoff
and the parameters in Table 1 for $\s8=0.95$. Filled circles show \wp\
for the same values of $M_1$ and $\mcut$, but an exponential cutoff,
with $\mmin$ adjusted to retain the galaxy density. The differences in
the calculated correlation functions are barely discernible, mostly
confined to the two-halo term at small scales, where the one-halo term
dominates \wp\ anyway. Figure \ref{hod_compare}b shows an example of
\clf\ for the soft and hard central cutoffs for $\s8=0.95$ at
$M_h=3\times 10^{13}$ \hmsol. The low luminosity regimes are
identical, but at high luminosities where central galaxies dominate,
the central cutoff produces a well-defined bump, which is smoothed out
in the exponential cutoff model. Panel (c) shows the mass-to-light
ratio as a function of halo mass for these two prescriptions. The
sharp cutoff model produces artificial jumps in the \ml\ function, but
the overall trend and, more importantly, the behavior at high halo
masses are the same for hard and soft cutoffs. We conclude that the
details of the central galaxy cutoff do not significantly affect the
mass-to-light ratio predictions.

\begin{figure}
\epsscale{1.0} 
\plotone{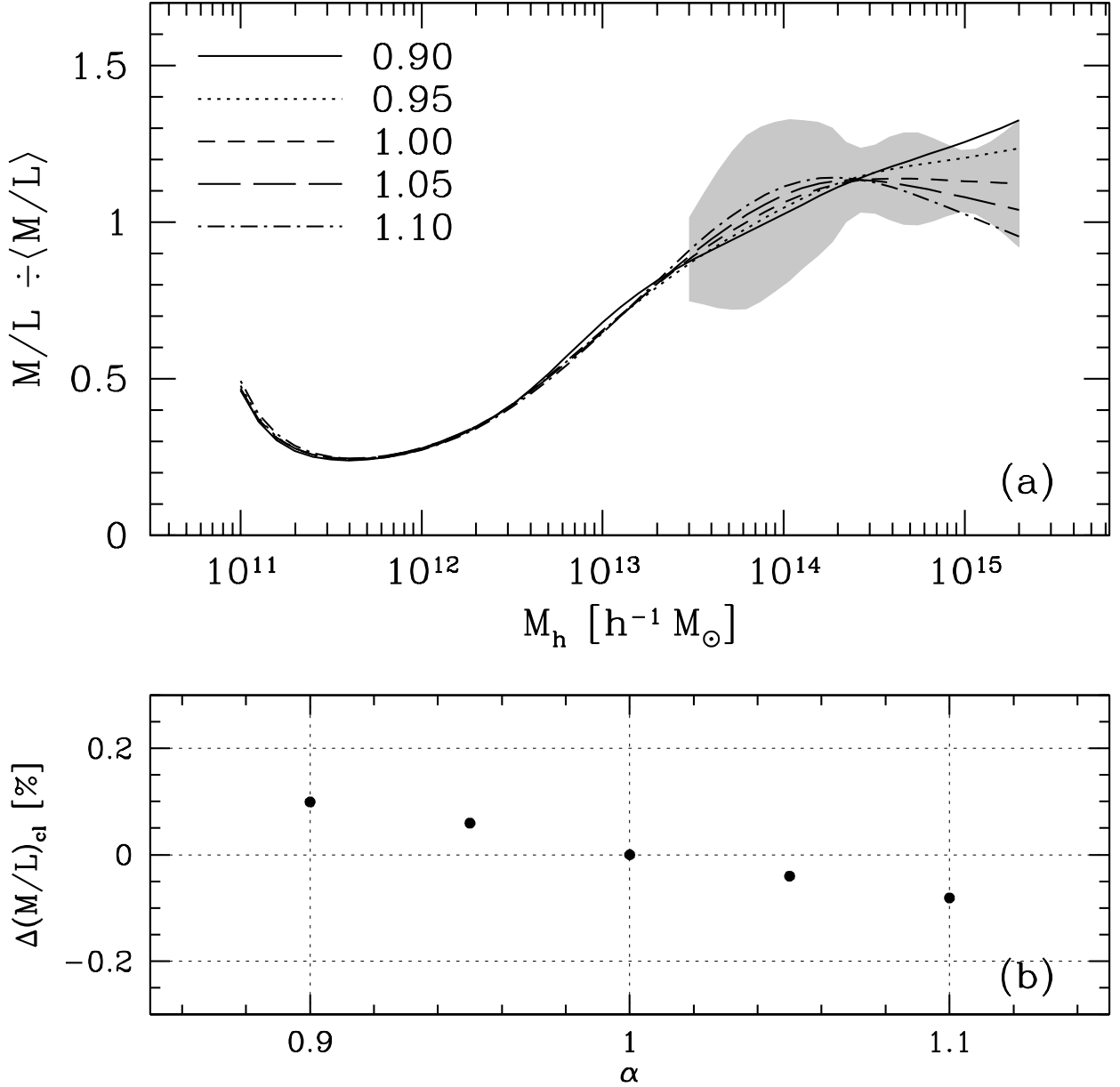}
\caption{ \label{vary_alpha} (a) \ml\ ratio as a function of halo mass
  for five values of $\alpha$: 0.9, 0.95, 1.0, 1.05 and 1.1. All
  calculations assume $\s8=0.9$. The shaded region is the full-width
  at 10\% maximum of the distribution of halo \ml\ inferred from a
  flexible HOD parameterization with no fixed value of $\alpha$. See
  text for further details. (b) The change in the mean cluster \ml\
  ratio as a function of $\alpha$ relative to $\alpha=1$. The mean is
  calculated for the same distribution of cluster masses as the CNOC
  cluster sample. }
\end{figure}

In Figure \ref{vary_alpha} we quantify the dependence of our results on
the value of $\alpha$. First, we refit \wp\ and calculate \ml\ for
$\s8=0.9$ with $\alpha=0.9$, 0.95, 1.05, and 1.1. The results are shown,
along with the fiducial results for $\alpha=1$, as the five curves in
Figure \ref{vary_alpha}a. The \ml\ ratios are nearly identical at group
masses and below. At $M \approx 2\times 10^{14}$ \hmsol\ the curves
begin to separate, with lower $\alpha$ resulting in higher \ml, and vice
versa, due to the different scalings of galaxy number with halo mass.

Our second method for testing the sensitivity to $\alpha$ uses a much
more flexible HOD parameterization in which the value of satellite
mean occupation is specified at $\log M = 12,$ 13, 14, and 15,
connected by a cubic spline and smoothly truncated at low masses (see
Figure~19 of \citealt{zehavi04b} for examples of similar fits).  With
this extra freedom, it is difficult to guarantee that $\nsat$ is a
monotonically increasing function of luminosity threshold, so we have
therefore fit the \cite{zehavi04b} measurements for samples in
absolute magnitude bins, rather than magnitude thresholds. In each
bin, we assume that $\ncen=1$ in a range $M_{\rm c,min}-M_{\rm c,max}$
and $\ncen=0$ elsewhere, equivalent to our sharp cutoff assumption for
luminosity threshold samples.  Using a Monte Carlo Markov Chain (e.g.,
\citealt{lewis02}), we identify between 17 and 42 models for each
luminosity bin that have $\Delta\chi^2 \leq 1$ relative to the best
fit model.  Using all $4.7\times 10^6$ possible combinations of these
models, we evaluate the $M/L$ ratios at the cluster mass scale.  The
distribution of $M/L$ values with respect to the mean at a given mass
has an approximately Gaussian core with a half-width of 11\% at 10\%
of the maximum.  At lower probabilities, there is a tail toward high
$M/L$ values that arises from HOD fits that are statistically
acceptable but physically implausible, with anomalously high halo
masses for central galaxies.  There is no corresponding tail to low
$M/L$ values, since if one puts too many galaxies in clusters the
correlation function is inevitably too high. The shaded region in
Figure \ref{vary_alpha}a shows the full-width at 10\% maximum as a
function of halo mass.  This test also accounts for statistical
uncertainty in the \wp\ measurements themselves.

Figure \ref{vary_alpha}b quantifies the change in the average cluster
\ml\ as a function of $\alpha$. The $y$-axis shows the percentage
difference in \mlcl, averaged over the halo masses of the CNOC cluster
sample (see \S 3.2 below), with respect to $\alpha=1.0$. For $\Delta
\alpha = 0.1$, $\Delta$\mlcl\ $\approx 8\%$, and for $\Delta \alpha =
0.05$, $\Delta$\mlcl\ $\approx 4\%$. Although the differences in the
\ml\ curves of panel (a) appear large at the highest halo mass, the
impact of these differences in moderated in panel (b) because clusters
above $10^{15}$ \hmsol\ are relatively rare. We conclude that for any
given cosmology, the existing \wp\ measurements constrain the average
cluster mass-to-light ratio, the quantity most relevant to our
conclusions, to within about 10\%. However, the detailed shape of the
\ml\ curve at $M_h > 3\times 10^{13}$ \hmsol\ depends on the assumed
form of the HOD.


\begin{deluxetable}{cccccccccccccccc}
\tablecolumns{13} 
\tablewidth{40pc} 
\tablecaption{HOD Parameters for the SDSS Galaxy Samples}
\tablehead{

\colhead{$M_r$} & 
\colhead{$\mmin$} & \colhead{$\mcut$} & \colhead{$M_1$} & \colhead{$\chi^2_{dof}$} &
\colhead{$\mmin$} & \colhead{$\mcut$} & \colhead{$M_1$} & \colhead{$\chi^2_{dof}$} &
\colhead{$\mmin$} & \colhead{$\mcut$} & \colhead{$M_1$} & \colhead{$\chi^2_{dof}$} \\

}
\startdata

\multicolumn{1}{c}{ } & 
\multicolumn{4}{c}{$\s8$=0.95} &
\multicolumn{4}{c}{$\s8$=0.9} &
\multicolumn{4}{c}{$\s8$=0.8} \\ \hline \\

-18.0 & 11.2 & 12.0 & 12.7 & 1.63 & 11.3 & 11.7 & 12.7 & 1.40 & 11.3 & 12.1 & 12.6 & 1.11 \\
-18.5 & 11.4 & 12.3 & 12.8 & 0.88 & 11.4 & 12.3 & 12.7 & 0.72 & 11.4 & 12.7 & 12.6 & 0.49 \\
-19.0 & 11.5 & 12.7 & 12.8 & 0.94 & 11.5 & 12.7 & 12.8 & 0.90 & 11.5 & 12.9 & 12.7 & 1.04 \\
-19.5 & 11.7 & 13.1 & 13.0 & 0.22 & 11.7 & 13.1 & 12.9 & 0.22 & 11.7 & 13.1 & 12.9 & 0.51 \\
-20.0 & 12.0 & 13.0 & 13.3 & 0.51 & 12.0 & 13.0 & 13.2 & 0.51 & 12.0 & 13.2 & 13.1 & 0.58 \\
-20.5 & 12.3 & 13.1 & 13.5 & 1.80 & 12.3 & 13.2 & 13.4 & 2.21 & 12.3 & 13.3 & 13.4 & 3.09 \\
-21.0 & 12.7 & 13.6 & 13.9 & 1.28 & 12.7 & 13.6 & 13.8 & 1.36 & 12.7 & 13.7 & 13.7 & 1.58 \\
-21.5 & 13.3 & 14.6 & 14.1 & 0.94 & 13.3 & 14.6 & 14.0 & 0.90 & 13.2 & 14.6 & 13.8 & 0.86 \\
-22.0 & 13.9 & 14.8 & 14.6 & 1.42 & 13.9 & 14.7 & 14.5 & 1.42 & 13.8 & 14.9 & 14.2 & 1.27 \\
 
\\
\hline
\\

\multicolumn{1}{c}{ } & 
\multicolumn{4}{c}{$\s8$=0.7} &
\multicolumn{4}{c}{$\s8$=0.6} \\ \hline \\

-18.0 & 11.3 & 12.1 & 12.5 & 0.96 & 11.3 & 12.1 & 12.4 & 0.94 \\
-18.5 & 11.4 & 12.6 & 12.6 & 0.43 & 11.4 & 12.5 & 12.5 & 0.61 \\
-19.0 & 11.5 & 12.8 & 12.7 & 1.43 & 11.5 & 12.7 & 12.6 & 2.11 \\
-19.5 & 11.7 & 13.0 & 12.8 & 1.17 & 11.7 & 12.9 & 12.7 & 1.98 \\
-20.0 & 12.0 & 13.1 & 13.0 & 0.70 & 12.0 & 13.2 & 12.9 & 0.91 \\
-20.5 & 12.3 & 13.3 & 13.3 & 3.97 & 12.2 & 13.3 & 13.1 & 5.02 \\
-21.0 & 12.7 & 13.6 & 13.6 & 1.86 & 12.6 & 13.7 & 13.4 & 2.31 \\
-21.5 & 13.2 & 14.5 & 13.6 & 0.88 & 13.1 & 14.5 & 13.4 & 1.00 \\
-22.0 & 13.7 & 14.9 & 13.8 & 0.97 & 13.6 & 14.6 & 13.8 & 0.74 \\

\enddata
\tablecomments{Values of $\mmin$,
$\mcut$, and $M_1$ are listed as $\log(M/h^{-1}$M$_\odot)$}
\end{deluxetable}


\section{Mass-to-Light Ratios of Large-Scale Structure}

\subsection{\ml\ vs. overdensity}

The numerical results in this and subsequent sections are calculated
using Tinker, Weinberg, \& Zheng's (2004) \lcdm\ simulations
(inflationary cold dark matter with a cosmological constant). These
simulations, consisting of a set of 5 realizations of $360^3$
particles evolved in a periodic volume 253 \hmpc\ on a side, have the
appropriate mass resolution for modeling the luminosity-dependent \xg\
of galaxies with $M_r\le -20$. Fainter samples have minimum halo
masses below the smallest resolved halos in our simulations, which are
identified using a friends-of-friends algorithm with a minimum of 30
particles. The dark matter power spectrum used to create the initial
conditions is identical to that assumed in the analytic HOD
calculations of the previous section. These simulations are run with
\gad\ (Springel, Yoshida, \& White 2000), with a force softening of 70
\hkpc, and the standard \gad\ timestep criterion $\eta=0.2$. The
simulations have $\om=0.1$, $\Lambda=0.9$, and $\s8=0.95$ at the last
output, and we use earlier outputs to represent $z=0$ results for lower
$\s8$, rescaling halo masses as necessary to represent different $\om$
(see \citealt{ztwb02} and \citealt{tinker04} for further
discussion). For the purposes of this paper, this rescaling approach
should yield nearly identical results to running separate simulations
for each $(\om,\s8)$ combination, since at fixed $\s8$ and $\plin$ the
value of $\om$ slightly alters the density profiles of halos but has
minimal impact of the halo masses and clustering.

Since \clf\ is known from the fits to \wp, it is a simple process to
create galaxy distributions from the dark matter halo populations in
the simulations. These galaxy distributions will match both the SDSS
clustering and luminosity function for all values of $\s8$. For each
halo above $\mmin$ (for the $M_r<-20$ sample), the central galaxy is
placed at the center of the halo, which we define as the location of
the most-bound dark matter particle, i.e., the particle with the most
negative potential energy calculated by a Newtonian summation over all
the particles in the halo.  The number of satellite galaxies is chosen
from a Poisson distribution with a mean given by equation
(\ref{e.nsat}), and the satellites are spatially distributed by random
sampling of the dark matter particles in the halo (other than the
most-bound particle). The luminosities of the galaxies are selected at
random from \clf, truncated at $M_r=-20$. Since central and
satellite galaxies have distinct contributions to \clf, in practice
they are treated separately when populating the N-body halos; central
galaxy luminosities are chosen only from their contribution to the
total \clf, while satellite luminosities are chosen from the satellite
galaxy portion of \clf.

The halo mass function in the simulations differs from the analytic
form of \cite{Jenkins01} at some masses. The discrepancy is within the
range shown in Jenkins \etal 's Figure 8, but at the high end, with
maximum difference of 10-15\% in number of halos at fixed mass. Using
our analytically derived HOD parameters, which assume the
\cite{Jenkins01} form, therefore leads to $\sim 7.5\%$ too many
galaxies (relative to the observed space density we are trying to match)
when we populate the simulations. We correct for this discrepancy when
populating the N-body simulations by changing $\mmin$ for each sample to
the value that yields the correct number of galaxies. This change is
moderate, $\lesssim 10\%$ in $\mmin$, but it is required in order for
the luminosity function in the simulations to accurately represent that
of the \cite{zehavi04b} samples.

A simple application of these galaxy distributions is to investigate
mass-to-light ratios as a function of overdensity in randomly placed
spheres, as done by B00 using their hydrodynamic simulation. Figure
\ref{ml_spheres} plots the mass-to-light ratio, relative to the mean
\ml\ of the box, as a function of dark matter density at three top-hat
smoothing scales. (In this way of plotting our data, values greater than
one would be considered ``anti-biased'' by the definition of B00, since
luminosity enters in the denominator). For each $\delta$, the average
light and mass are calculated separately and then used to calculate
\ml. This prevents divergence in spheres with no galaxies, but in
practice it does not noticeably change the curves. Panel (a) shows the
results for a top-hat smoothing scale of 1.5 \hmpc, the scale used by
\cite{abell58} to define galaxy clusters, which is close to the virial
radius of $\sim 10^{15}$ \hmsol\ halos. The curves shoot up at
$\delta\lesssim0$, since at these overdensities there are no halos above
$\mmin$ and therefore no galaxies, so $\plum$ approaches zero faster
than $\rho_m$. The exact value of $\delta$ at which this upturn occurs
depends on the adopted luminosity threshold, since fainter galaxies
occupy lower mass halos, and on the smoothing radius, since the
probability of finding no halos above $\mmin$ drops with increasing
radius. The sharpness of the upturn depends on the sharpness of the
$\navg$ threshold at $\mmin$, but the qualitative behavior at
$\delta\lesssim0$ is a generic prediction of any model with a minimum
halo mass. In the range $0 \lesssim \delta \lesssim 10$, all the curves
dip below the mean due to low-mass halos that host a central
galaxy. Overdensities of $\delta \gtrsim 100$ begin to represent
virialized, cluster-like structures. At this $\delta$, the different
$\s8$ models spread out, with high-$\s8$ having \ml\ ratios above or
equal to the mean and low-$\s8$ models having \ml\ as low as half the
global mean. Panels (b) and (c) show the same results but for smoothing
scales of 5 and 10 \hmpc. For both of these smoothing scales, the
different $\s8$ models diverge at the highest overdensities, with low
$\s8$ corresponding to \ml\ ratios below the mean of simulation. At the
highest overdensities, most of the \ml\ curves flatten to a roughly
constant value, though for low $\s8$ and large smoothing scale the \ml\
ratio is a declining function of $\delta$ in this regime. More
importantly, the values of \ml\ at high $\delta$ are nearly independent
of the smoothing scale for a given model, even when they are far
from the universal value.

The implications of Figure \ref{ml_spheres} are clear. The existence
of a plateau in \ml, either as a function of $\delta$ at fixed
smoothing length or as a function of smoothing scale at fixed
$\delta$, cannot be taken as evidence that one has measured the
universal value of \ml. Simply measuring \ml\ in a large volume does
not guarantee convergence to the universal \mlmean\ if the
measurements are made only in dense regions, where galaxies may be
over- or under-represented, depending on the cosmological model.
Given the level of galaxy clustering in the SDSS, the plateau in \ml,
for galaxies brighter than $M_r=-20$, should be similar to the
universal mean if $\s8=0.9$, above the universal mean if $\s8>0.9$,
and below the universal mean if $\s8<0.9$. As we will demonstrate in
the next section, the value of $\s8$ for which the plateau is equal to
the universal value depends on the bias of the galaxy sample under
consideration, so for fainter luminosity thresholds the ``unbiased''
value of $\s8$ is lower.

\subsection{\ml\ of Clusters}

\begin{figure}
\epsscale{1.0} 
\plotone{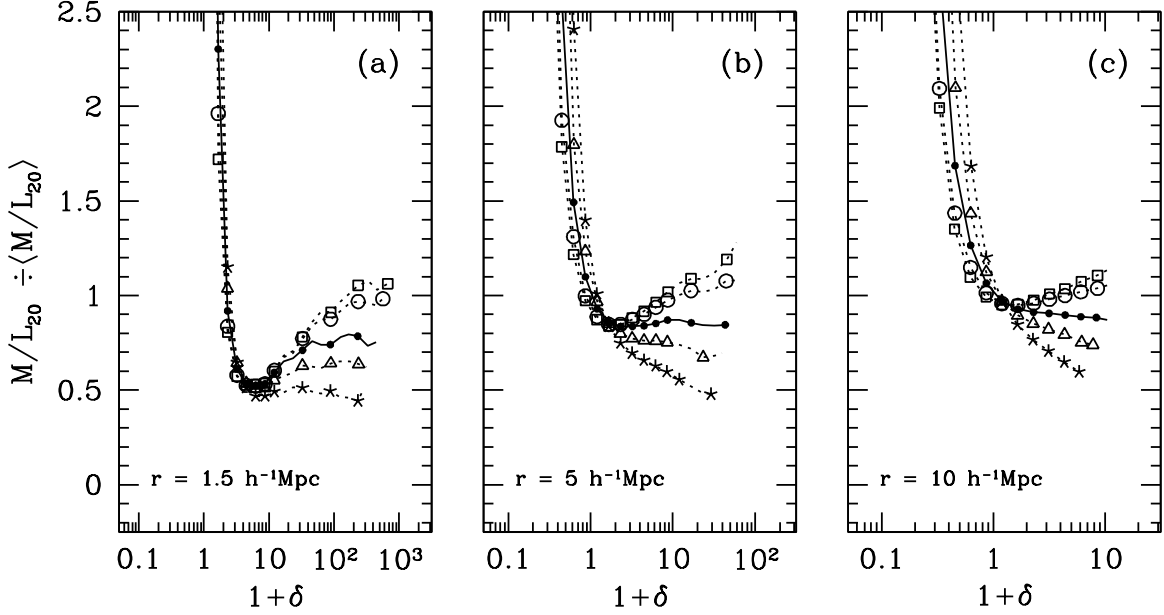}
\caption{ \label{ml_spheres} The mean \ml\ ratio in spheres of radius
  $r$ as a function of the enclosed overdensity $1+\delta$, for
  $r=1.5$, 5.0, and 10.0 \hmpc\ (panels a, b, and c,
  respectively). Results are calculated from the numerical simulations
  and refer to luminosities of galaxies with $M_r<-20$. Different
  symbols represent different $\s8$ values with the same coding as
  Figure \ref{clf}; at high $\delta$, the $\s8=0.6$ curve is always
  lowest.}
\end{figure}

\begin{figure}
\epsscale{1.0} 
\plotone{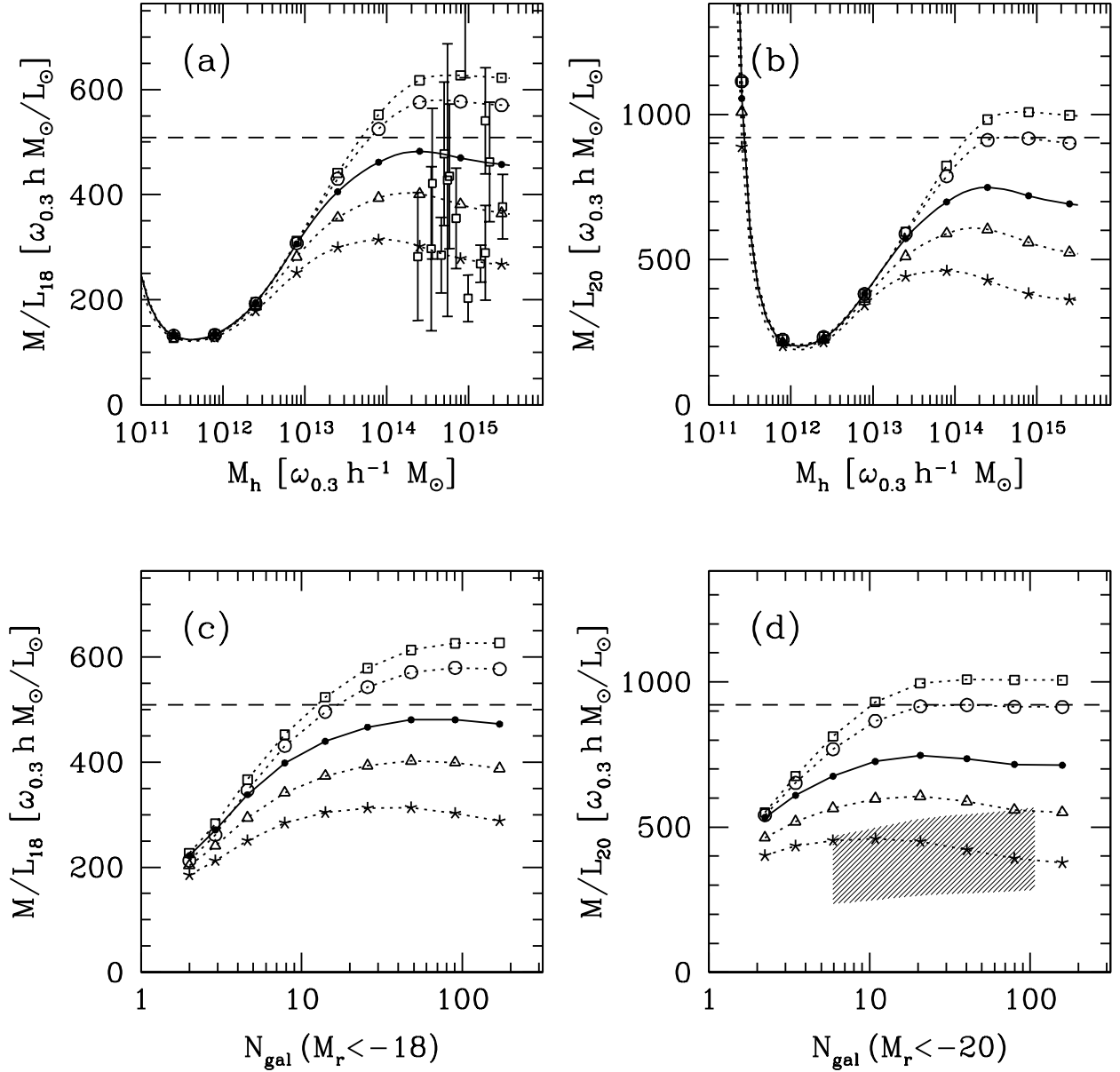}
\caption{ \label{ml_halo} Mass-to-light ratio ($r$-band) as a function
  of halo mass (top panels) or richness (bottom panels). Luminosities
  are for galaxies brighter than $M_r=-18$ (left) and $M_r=-20$
  (right). From bottom to top, curves represent $\s8=0.6, 0.7, 0.8,
  0.9, 0.95$. Dashed horizontal lines represent the mean \ml\ of the
  universe. In panel (a), open squares with error bars are the CNOC
  data of \cite{carlberg96}. In panel (d), the shaded region
  represents the \ml\ ratio of SDSS clusters based on
  \cite{bahcall03a}.}
\end{figure}

The \ml\ ratios shown in Figure \ref{ml_spheres}, averaged over
top-hat spheres of constant $\delta$, are not directly observable. In
Figure \ref{ml_halo}, we make quantitative predictions for cluster
sized halos that can be compared to observational data. For each
$\s8$, equation (\ref{e.ml_halo}) is used to calculate \ml\ as a
function of halo mass. Note that this calculation does not use
simulations, so it can include galaxies down to $M_r = -18$. We
present predictions for galaxies with $M_r<-18$ and for galaxies with
$M_r<-20$, and refer to these mass-to-light ratios as \mla\ and \mlb,
respectively.

Figure \ref{ml_halo}a plots \mla, calculated using the soft central
cutoff, against halo mass for all values of $\s8$. Note that halo
masses are proportional to $\om$ for fixed $\s8$, so we list masses in
\whmsol\ and mass-to-light ratios in \whml. The \ml\ vs. $M_h$ curves
derived from our modeling of \wp\ are similar in form to the
parameterized function used by \cite{Yang03} in their CLF modeling,
and to the results of semi-analytic modeling by \cite{benson00}. All
the curves in panel (a) show a clear minimum at $\sim 4\times 10^{11}$
\whmsol. Below this minimum halo mass, \mla\ increases rapidly, as it
must to match the observed galaxy luminosity function to the steeper,
low mass end of the halo mass function (see, e.g.,
\citealt{Yang03}). At higher halo masses, \ml\ rises less rapidly and
eventually reaches a maximum between $1-5 \times 10^{14}$ \whmsol,
with higher values of $\s8$ reaching maximum \mla\ at higher
masses. At still higher masses, \mla\ gradually declines, with the
results for $\s8=0.6$ falling by $\sim 20\%$ between $10^{14}$ \hmsol\
and $10^{15}$ \hmsol.

The mean mass-to-light ratio of the universe, for galaxies brighter than
$M_r<-18$, can be calculated by integrating the Blanton \etal\ (2003)
Schechter function fit to the observed $r$-band luminosity function of
SDSS galaxies. This gives a luminosity density of $1.63\times 10^8 h$
L$_\odot$ Mpc$^{-3}$, using an $r$-band absolute magnitude of the sun of
4.76, also taken from Blanton \etal\ (2003). Dividing the matter
density, $\rho_m = \om\times \pcrit=2.78\times
10^{11}\,h^2\om\mbox{M}_\odot\,\mbox{Mpc}^{-3}$, by this luminosity
density gives the mean $r$-band mass-to-light ratio of the universe, 509
\whml.\footnote{To be precise, note that the \cite{blanton03} luminosity
function is calculated in $^{0.1}r$, the SDSS $r$-band redshifted to
$z=0.1$, and that this shifted bandpass is also used to define the
luminosity threshold samples in \cite{zehavi04b}.} A similar value of
521 \whml\ is obtained by simply taking the number density in each
magnitude bin listed in \cite{zehavi04b} Table 2 and using a discrete
sum over all nine bins. The results in Figure \ref{ml_halo}a are similar
to those of Figure \ref{ml_spheres}a, but with a different model being
unbiased with respect to \mlmean; the halos above $10^{14}$ \hmsol\ for
the $\s8=0.8$ model have \mla $\approx 490$ \whml, which is very close
to the cosmic mean value. For $\s8$ above and below 0.8, the cluster
\mla\ ratios are above and below this mean.

Also plotted in Figure \ref{ml_halo}a are data from the Carlberg
\etal\ (1996) analysis of clusters from the CNOC survey, with \ml\ in
Gunn $r$-band calculated using mass estimates from the virial
theorem. To properly compare our calculations to their results we have
taken the values listed in their Table 4, which have been extrapolated
to include all luminosities below their magnitude limit of $-18.5$ (a
$\sim 15\%$ correction), and removed most of this correction up to our
magnitude limit of $-18.0$ using the Schechter function parameters
listed in their paper. This correction increases \ml\ ratios by 6\%
relative to their stated values. We have also included a moderate
correction for redshift evolution, since the mean redshift of the CNOC
data is $z=0.3$, while the SDSS data are centered on $z=0.1$. Using
the approximate correction factor of $10^{0.15 \Delta z}$
(Carlberg \etal\ 1996) we applied an additional 7\% reduction of the
luminosities. We ignore the slight differences between Gunn $r$ and
SDSS $^{0.1}r$, which should be largely removed by the correction to
solar units. The error-weighted mean of the Carlberg \etal\ data is
$359 \pm 32$ \hml.\footnote{This mean value does not include the two
clusters from the sample that show strong binarity. The error is
computed by the bootstrap method.}  Inserting this value into equation
(\ref{e.ml_omega}), i.e., assuming that it represents \mlmean, gives
$\om\approx0.2$. However, Figure \ref{ml_halo}a demonstrates that the
CNOC results are consistent with $\om=0.3$ if $\s8\approx 0.65$, since
the cluster \ml\ is then below the universal value.

A number of authors have reported a trend of increasing \ml\ with
cluster mass (e.g., \citealt{bahcall02, lin04, popesso04}), a result
seemingly in conflict with the claimed plateau of \ml\ at high mass and
with our results here. In the observational data and in our \ml\ curves,
there is a significant increase in \ml\ from group masses ($\sim
10^{13}$\hmsol) to the cluster mass regime. At $M \gtrsim 10^{14}$
\hmsol, the observational data of \cite{bahcall02} and \cite{popesso04}
are consistent with a horizontal line. \cite{lin04} find a positive
slope well into the cluster mass regime, but \cite{kochanek03},
analyzing a similar sample with different methods, find a {\it decrease}
in \ml\ with cluster mass, suggesting a significant systematic
uncertainty in the detailed behavior at high masses. Our predicted \ml\
curves imply that a single relation between $M$ and \ml\ is a poor
approximation for any samples that extend below $\sim 2\times 10^{14}$
\hmsol. Because of the observational uncertainties in the \ml\ trend and
the dependence of the predicted trend on the assumed value of $\alpha$
(see Figure \ref{vary_alpha}a), we do not use this trend to draw
cosmological conclusions. Instead, we use only the mean \ml\ in the
cluster mass regime, a quantity that is more robust observationally and
theoretically (see Figure
\ref{vary_alpha}b).

Panel (b) of Figure \ref{ml_halo} shows the same calculation as panel
(a), but now we only consider galaxies brighter than $M_r=-20$. Using
the \cite{blanton03} luminosity function, the mean \ml\ of the universe
for this magnitude threshold is 923 \whml. The predicted \ml\ ratios
still separate at high mass in the same proportions as in panel (a), but
now the $\s8$ for which \mlcl $\approx$ \mlmean\ is 0.9 instead of 0.8
for \mla, as with the numerical results presented in Figure
\ref{ml_spheres}.

Figure \ref{ml_halo}c plots \mla\ against galaxy
multiplicity rather than halo mass. We compute this relation by
integrating over the halo mass function to calculate the contribution
of each halo mass to the abundance of clusters at a given $\ngal$,
i.e.

\begin{equation}
\label{e.xfunc}
M/L_{18}(N_{\mbox{\scriptsize gal}}) = \frac{ \int P(\ngal|M)\, (M/L)_M\, \frac{dn}{dM}\, dM }
 { \int P(\ngal|M)\, \frac{dn}{dM}\, dM },
\end{equation}

\noindent where all values with a subscript $M$ are values at a given
halos mass and $dn/dM$ is the \cite{Jenkins01} halo mass function. The
probability $P(\ngal|M)=\nsat^{\ngal-1}\mbox{e}^{-\nsat}/{(\ngal-1)!}$
for $M\ge\mmin$, since the distribution of satellite numbers in Poisson,
and a halo with a satellite has a central galaxy by definition. Because
of the exponential cutoff at the high-mass end of the halo mass
function, there is asymmetric scatter into a given value of $\ngal$;
Poisson fluctuations around $\navg$ cause more halos of lower mass to
scatter to high $\ngal$ than vice versa. This effect flattens out the
curves relative to those of panel (a), but the asymptotic behavior at
high multiplicity is similar.

Figure \ref{ml_halo}d shows the same calculation for \mlb\ rather than
\mla. At the same halo mass, the number of galaxies with $M_r<-18$ is
roughly five times the number with $M_r<-20$, which is why the \ml\
curves reach their asymptotic values at $\ngal \sim 20$ rather than
$\sim 100$. As in panel (b), the asymptotic value of \ml\ for $\s8=0.9$
is closest to the mean value.

Figure \ref{ml_halo}d can be compared to the results from
\cite{bahcall03a} analysis of clusters in the SDSS Early Data
Release. We consider the cluster sample identified by the maxBCG
method (see, e.g. \citealt{hansen04}), which characterizes richness by
$\nbcg$, the number of galaxies close to the brightest galaxy in a
restricted region of color-magnitude space. \cite{bahcall03a} report
scaling relations

\begin{equation}
\label{e.NL}
L^r_{0.6} \,(10^{10} \, \mbox{L}_\odot) = 1.42 \nbcg,
\end{equation}

\begin{equation}
\label{e.Nsigv}
\sigma_v \,(\mbox{km s}^{-1}) = 93 \nbcg^{0.56},
\end{equation}

\noindent for luminosity and velocity dispersion as a function of
$\nbcg$. We have used the Schechter function parameters given in
Bahcall \etal\ (2003a) to correct equation (\ref{e.NL}) from their
observed limit of $M_r=-19.8$ to our threshold of $M_r=-20$. The
subscript 0.6 indicates that the luminosity is the total value within
0.6 \hmpc\ of the cluster center, the radius at which all the cluster
attributes are calculated by Bahcall \etal\ (2003a). The velocity
dispersion can be converted to mass through a relation calibrated on
gravitational lensing measurements of cluster masses (Bahcall \etal\
2003b):

\begin{equation}
\label{e.v2mass}
M_h (r<0.6 \,h^{-1} \mbox{Mpc}) = 3.28\times 10^{9} \sigma_v^{1.67} 
k_\delta \,\,h^{-1} \mbox{M}_\odot,
\end{equation}

\noindent where \kd\ is a small correction factor that depends on the
mean overdensity of the halo within the defined radius (see Evrard,
Metzler, \& Navarro 1996). Since \kd\ requires knowledge of the halo
mass, we solve equation (\ref{e.v2mass}) by iteration, then combine
with equation (\ref{e.NL}) to obtain \mlb\ for $8\le \nbcg \le 40$,
the range over which the scaling functions are valid. We convert from
$\nbcg$ to the number of galaxies above $M_r=-20$ using $\ngal =
0.14\,\nbcg^{1.8}$, an approximate scaling determined from the
dependence of the mean cluster luminosity function on $\nbcg$
(\citealt{wechsler04}).

The shaded region in Figure \ref{ml_halo}d encloses the mean relation
derived from equations (\ref{e.NL})$-$(\ref{e.v2mass}) and is bounded by
the 2-sigma errors on the scaling coefficient in equation
(\ref{e.NL}). To make this comparison, we have also assumed that \mlb\
does not vary from 0.6 \hmpc\ to the edge of the cluster. In our
simulations, we find a modest increase in \ml\ from 0.6 to 1.5 \hmpc\ of
about 20\%, due to the bright central galaxy in each halo. The large
contribution of the central galaxy to the overall luminosity of the
cluster is also seen in the cluster luminosity functions of
\cite{hansen04}. However, the trend with radius is much smaller than the
statistical errors on the scaling relations themselves.

There are significant systematic uncertainties in
our comparison because we combine scaling relations that have large
individual uncertainties and intrinsic scatter. Future \ml\
measurements for larger SDSS cluster samples will enable more direct
comparisons. For the current data, the mean relation plotted in Figure
\ref{ml_halo}b seems consistent with the \cite{carlberg96}
results. For a universe with $\om=0.3$, the mean relation is
consistent with $\s8=0.6$, while a lower value of $\om\approx 0.12$ is
required to match the observations for $\s8=0.9$.

\begin{figure}
\epsscale{1.0}
\plotone{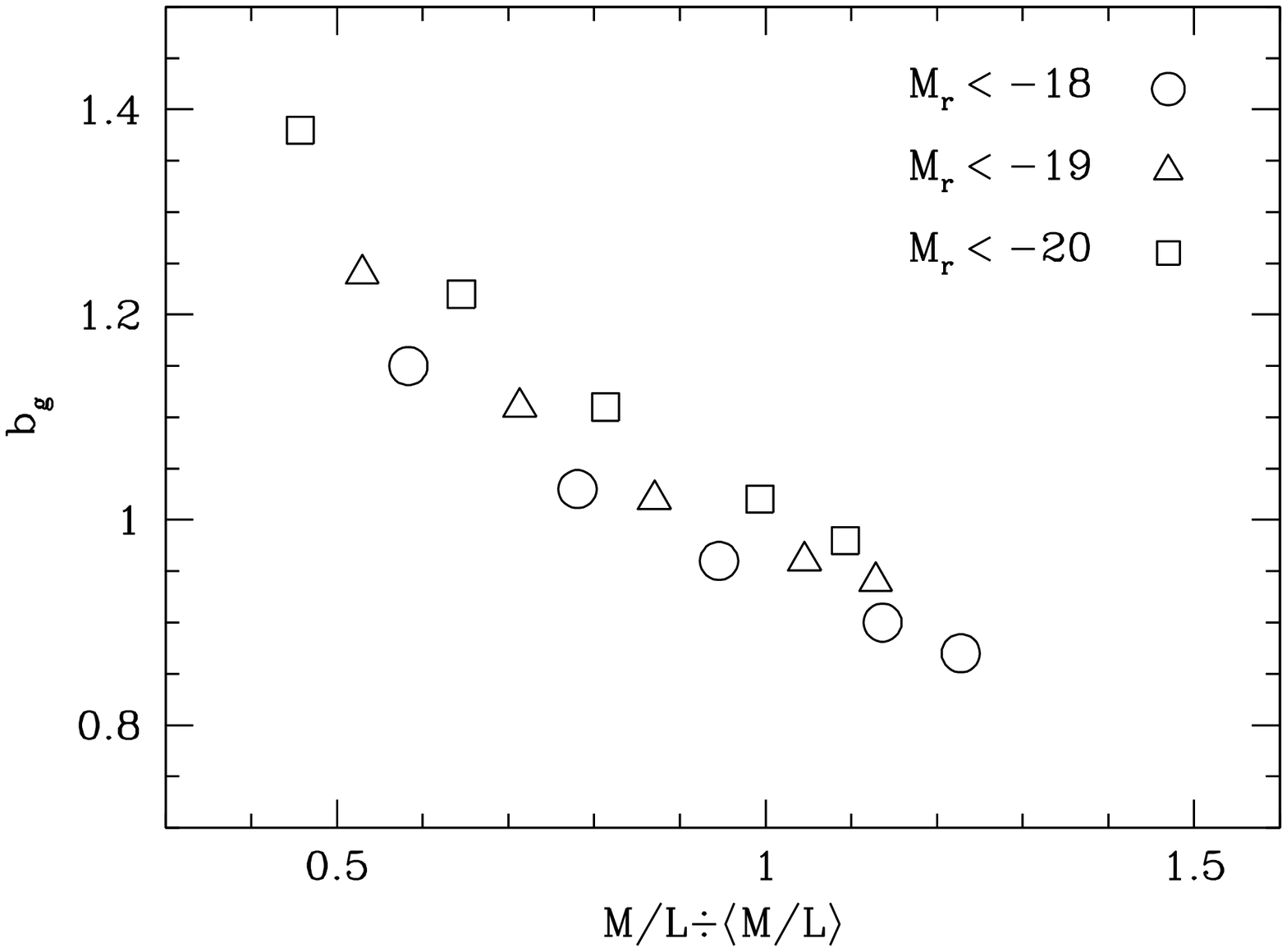}
\caption{ \label{bias} Galaxy bias, calculated for the SDSS samples
  from our HOD models, as a function of mass-to-light ratio at $M=5\times
  10^{14}$ \whmsol, relative to \mlmean. The different plot symbols
  represent three different galaxy magnitude thresholds, $M_r=-18,
  -19$, and $-20$. For each sample, the data points represent the five
  different $\s8$ values used. The data always follow the monotonic
  trend of increasing $\s8$ for decreasing $b_g$ (e.g., $\s8=0.6$ at
  the far left and $\s8=0.95$ at the far right). }
\end{figure}

At high halo masses, the values of \mla\ and \mlb\ are close to the
universal values for $\s8=0.8$ and $\s8=0.9$, respectively. The
difference reflects the higher amplitude correlation function of more
luminous galaxies. For a given value of $\s8$, reproducing this trend
requires putting a larger fraction of the more luminous galaxies in
the strongly clustered, high mass halos. On large scales, the galaxy
correlation function is $\xi_{gg} = b^2_g\xi_{mm}$, where the galaxy
bias factor $b_g$ is a number-weighted average of the halo bias factor
$b_h(M)$:

\begin{equation}
\label{e.bg}
b_g = \ngavg^{-1}\int_{\mmin}^\infty \,b_h(M)\, \navg \, \frac{dn}{dM}\,dM,
\end{equation}

\noindent where $\ngavg$ is the mean galaxy number density. Figure
\ref{bias} plots $b_g$ for luminosity thresholds $M_r=-18, -19$, and
$-20$, against the ratio \mlclnormp, where \mlclp\ is evaluated at
$M=5\times 10^{14}$ \whmsol\ and \mlmean\ is the universal
mass-to-light ratio. For a given luminosity threshold, $b_g$ and
\mlclp\ are decreasing and increasing functions of $\s8$,
respectively, since matching the observed galaxy correlation function
requires a lower bias factor for higher $\s8$, and a lower bias factor
implies a smaller fraction of galaxies in high mass halos. At fixed
$\s8$, $b_g$ must be larger for high luminosity thresholds, and
\mlclnormp\ is correspondingly lower. In practice, we find in Figure
\ref{bias} that $(M/L)_{\rm cl} \approx \langle M/L\rangle$ for the
value of $\s8$ that has $b_g \approx 1$. There is no reason this must
be exactly true, but our results are well captured by a simple rule of
thumb: the cluster mass-to-light ratios for a given luminosity
threshold and $\s8$ are below the universal value if the large scale
galaxy correlation function is positively biased, above the universal
value if the correlation function is anti-biased, and equal to the
universal value if the correlation function is unbiased.

\subsection{\ml\ in cluster infall regions}

Attempts to measure \ml\ of cosmic structure often focus on galaxy
clusters, since their masses can be estimated by the virial theorem, by
more general dynamical models (e.g., \citealt{carlberg97}), by
modeling their X-ray emission, or by weak gravitational lensing.
Outside the virial region, there is still matter that is
gravitationally bound to the cluster, but it is not in dynamical
equilibrium, so the above methods (with the exception of weak lensing)
are inapplicable.  At the boundary of the infall region, where
peculiar velocity cancels Hubble flow, the galaxy phase space density
becomes infinite, creating caustic-like features in redshift
space. The amplitude of these caustics is a measure of the escape
velocity of the system (Kaiser 1987; Diaferio \& Geller 1997; Diaferio
1999). The goal of the CAIRNS survey (Rines \etal\ 2003, 2004) was to
identify these features and thereby measure the mass profile of
clusters out to $r\sim 10$ \hmpc.

To compare our predictions to the results of the CAIRNS survey, we
identify cluster-mass halos ($M\ge3\times 10^{14}$ \hmsol) in our
N-body simulations and calculate the average \ml\ ratio as a function
of radius. Since the CAIRNS sample is taken from 2MASS $K$-band data,
the \ml\ ratios need to be converted to $r$-band. The total
luminosities listed in Table 8 of \cite{rines04} are in solar $K$-band
units, corrected for incompleteness. Using $r-K$ colors of 1.43 for
the sun and 2.81 for elliptical galaxies (Pahre 1999), we multiply
$(M/L)_{K}$ by $10^{0.4\,(2.81-1.43)}=3.57$ to get $(M/L)_{r}$. We
have multiplied each \ml\ ratio by 1.7 to remove the luminosity from
galaxies fainter than $M_K=-22.81$ for a proper comparison to our
$M_r<-20$ predictions.

\begin{figure}
\epsscale{1.0}
\plotone{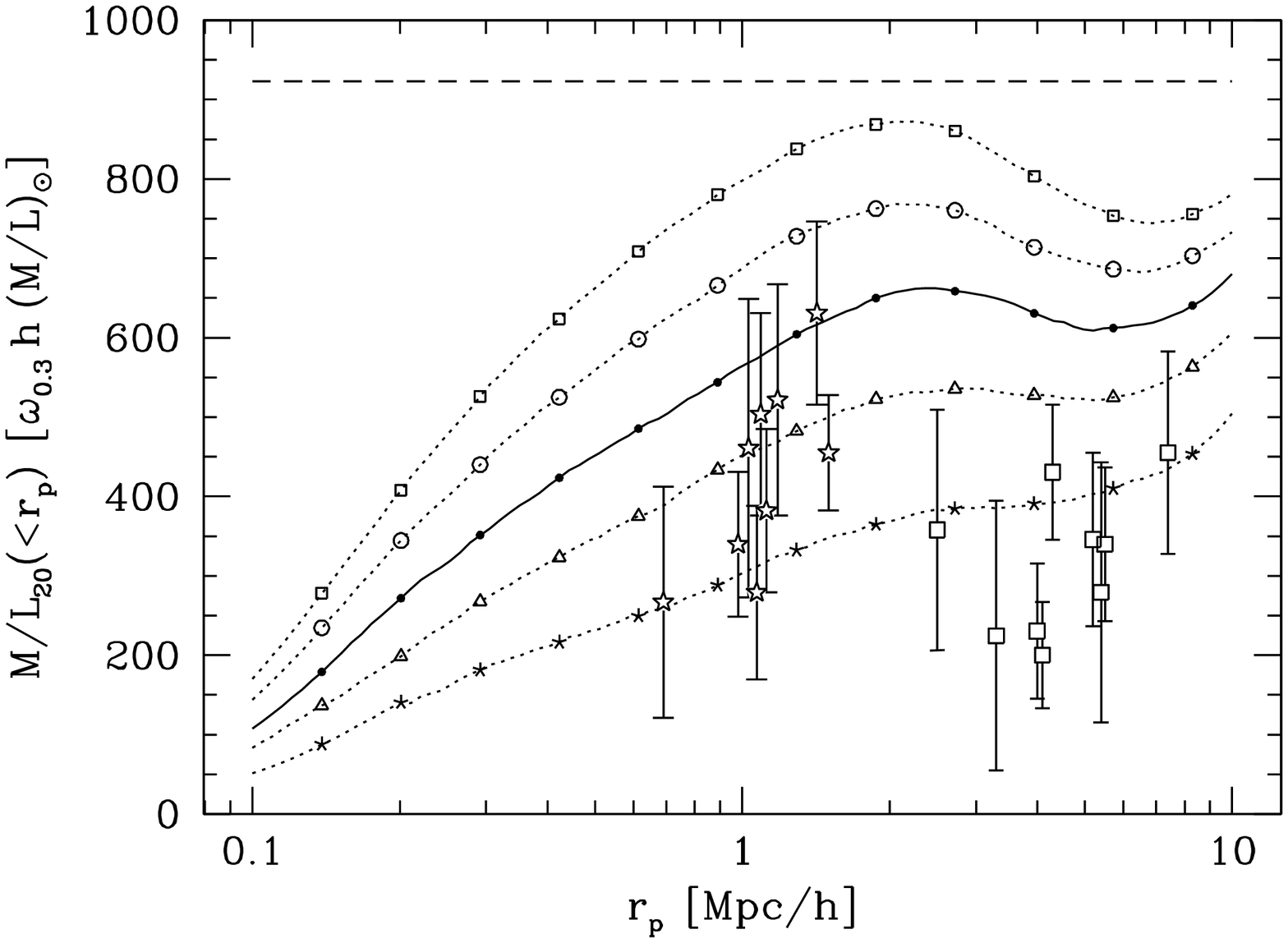}
\caption{ \label{ml_cluster} Mean \ml\ ratios of clusters (with $M>3
  \times 10^{14}$ \hmsol) as a function of projected separation from
  the cluster center, calculated from the numerical simulations. from
  bottom to top, the curves represent $\s8 = 0.6, 0.7, 0.8, 0.9,
  0.95.$ Following \cite{rines04}, we calculate masses in spheres and
  luminosities (for $M_r\ge -20$) in truncated cylinders, which
  depresses \ml\ values below those shown in Figures \ref{ml_halo}b
  and \ref{ml_halo}d. Stars and open squares represent the \ml\
  estimates of \cite{rines04} at $r_{200}$ and cluster turnaround
  radii, respectively, converted from $K$-band to $r$-band as
  described in the text. The dashed horizontal line shows the mean
  \mlb\ of the box. }
\end{figure}

The CAIRNS \ml\ ratios have a specific geometry: the total light is
in cylinders and the mass is in spheres. We calculate \mlb\ ratios
from our N-body simulations in the same way: centering on the
most-bound particle of each halo with mass larger than $3\times
10^{14}$ \hmsol, the light at each projected radius $r_p$ is
calculated within a cylinder that is extended 10 \hmpc\ in either
direction from the cluster center, using the $z$-axis of the box as
the line of sight. The total mass in dark matter particles is
calculated in spheres of radius $r_p$. The results for all $\s8$
values are shown in Figure \ref{ml_cluster}. The ratio of these
different geometries lowers \mlb, since the volume of the cylinder is
larger than that of the sphere. In this spheres-on-cylinders
calculation, the \mlb\ values at $3 < r_p < 10$ \hmpc\ are lower than
the mean, in contrast to the previous results. In tests that use
equal cylinders for both light and mass, the \mlb\ values, relative to
the mean of the box, appear much as they did in Figure \ref{ml_halo},
with the $\s8=0.9$ model lying closest to the universal \ml\ at large
$r_p$.

Points with error bars represent the CAIRNS measurements at $r_{200}$
(stars) and cluster turnaround radii (squares), taken from Table 8 of
\cite{rines04} and converted to $r$-band as described above. Here
$r_{200}$ is the radius at which the overdensity of the cluster is 200
times the {\it critical} density (666 times the mean density assuming
$\om=0.3$), which is close to 1 \hmpc\ for all the clusters in the
CAIRNS sample. The error-weighted mean of the $r_{200}$ results is
$424\pm 37$ \hml. For $\om=0.3$, this value for \mlclb\ implies
$\s8\approx 0.7$, similar to the CNOC results and the SDSS results
shown in Figure \ref{ml_halo}. For $\s8\approx 0.9$, a value of
$\om\approx 0.18$ is required to bring the numerical results into
agreement with the data. The error-weighted mean of the \mlb\ values
from the cluster+infall regions is significantly lower, $313\pm 32$
\hml. For $\s8=0.9$, this ratio implies $\om\approx 0.14$.

\begin{figure}
\epsscale{1.0}
\plotone{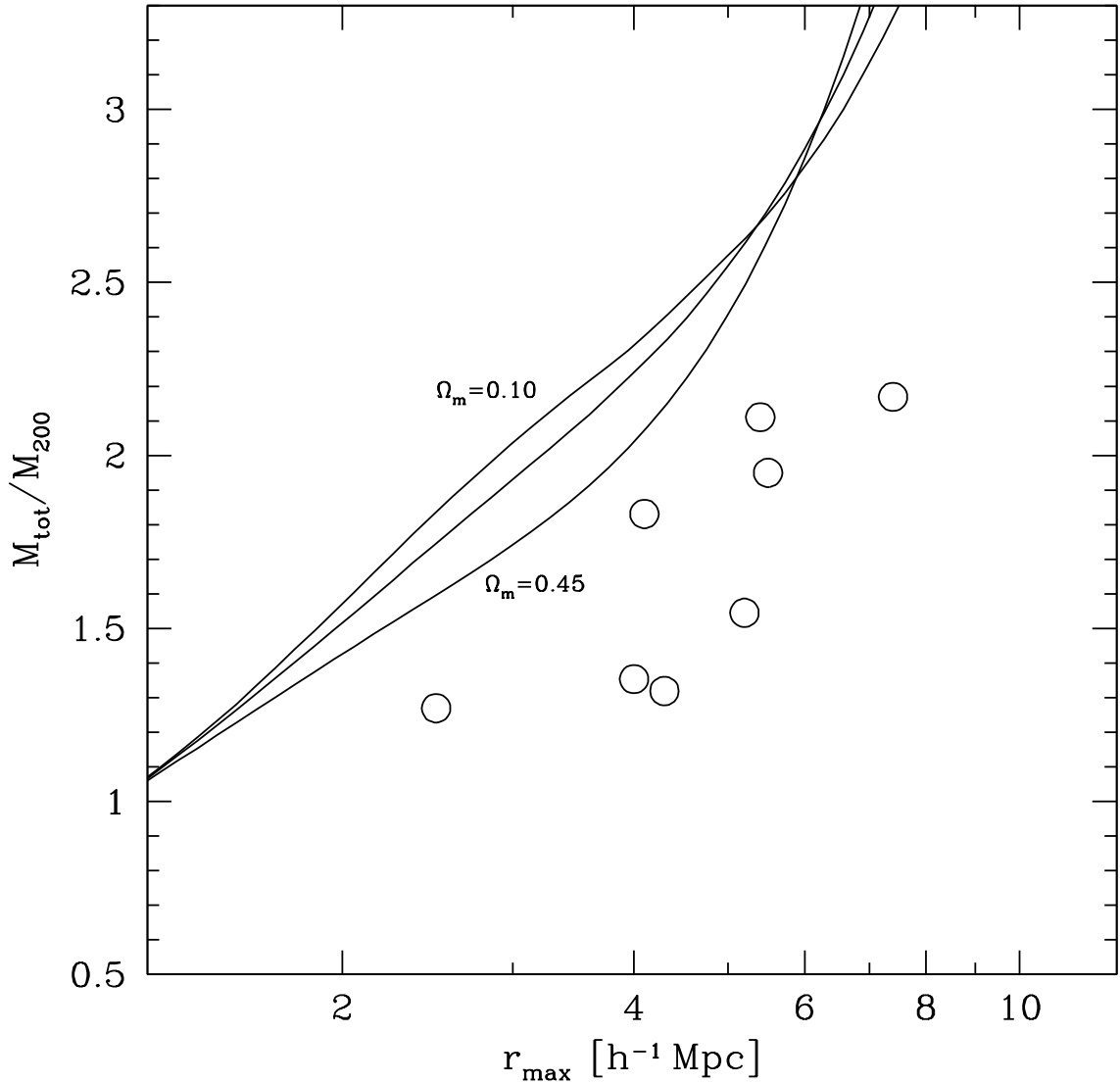}
\caption{ \label{mass_ratio} Ratio of the total mass within $\rmax$ to
  the mass within $r_{200}$ from the \cite{rines04} data (open
  circles) and our simulations (curves). The predictions depend
  slightly on the value of $\om$ assumed in calculating cluster
  masses, since we consider only halos with $M\ge 3 \times 10^{14}$
  \hmsol. The curves represent $\om=0.1, 0.3$, and $0.45$, and are
  calculated for $\s8=0.8$; results for other $\s8$ are similar.}
\end{figure}

The discrepant conclusions between the virial and infall mass
estimates suggest that the mass profiles inferred by \cite{rines04}
are steeper than those predicted by our simulations. To investigate
this point further, we plot in Figure \ref{mass_ratio} the ratio of
the mass $M_{\rm tot}$ within the turnaround radius $\rmax$ to the
mass $M_{200}$ within $r_{200}$, as a function of $\rmax$. The
observationally inferred mass ratios lie in the range 1.2 to 2.2, with
a trend of larger mass ratios for increasing $\rmax$. Solid curves
show the results for our simulated clusters. While $M_{\rm tot}$ and
$M_{200}$ in our simulations both scale with $\om$, there is still a
small dependence on $\om$ because we choose clusters above $3\times
10^{14}$ \hmsol, and therefore select a different sample for different
$\om$. Note that the observational data points are independent of the
galaxy luminosity measurements and the theoretical curves are
independent of our galaxy bias models, since there ratios refer to
mass alone. The theoretical curves lie above the data points by $\sim
50\%$, similar to the difference between the average virial and infall
mass-to-light ratios. Our best guess is that the caustic method
systematically underestimates the infall masses by $\sim 50\%$, but it
is of course possible that the observationally inferred mass profiles
are correct and conflict with the generic predictions of the \lcdm\
model.


\section{Implications and Outlook}

We have examined mass-to-light ratios of large scale structure in
cosmological models that are constrained to match observed real space
galaxy clustering.  Specifically, we consider models in which the
shape of the linear matter power spectrum is held fixed and the galaxy
halo occupation distribution is adjusted to reproduce Zehavi et al.'s
(\citeyear{zehavi04b}) measurements of the projected galaxy
correlation function.  For power spectrum normalization $\sigma_8$ in
the range $0.6-0.95$, we are able to find HOD parameters that yield
acceptable fits to the observed \wp, even with a restricted,
3-parameter HOD prescription.  For each value of $\sigma_8$, the $M/L$
ratios in high overdensity regions are approximately independent of
top-hat smoothing scale in the range $1.5-10$\hmpc, and the $M/L$
ratios of virialized halos climb from a minimum at $M_h \sim {\rm
several}\times 10^{11}$\hml\ to an approximately flat plateau in the
cluster mass regime.  However, this plateau only corresponds to the
true universal $M/L$ for a particular choice of $\sigma_8$, the one
for which the large scale galaxy correlation function is unbiased.
One therefore cannot take the existence of a plateau in $M/L$ as a
function of scale or of halo mass as evidence that one has measured
the universal $M/L$.  Estimates of $\Omega_m$ that multiply cluster
mass-to-light ratios by the observed luminosity density make the
implicit assumption that the galaxy distribution is unbiased.

Given the SDSS clustering measurements, we expect cluster mass-to-light
ratios to be representative of the universal value for $\sigma_8 \approx 0.8$
if one is considering galaxies with $M_r \leq -18$, or for
$\sigma_8 \approx 0.9$ if one is considering galaxies with $M_r \leq -20$.
For lower $\sigma_8$, cluster $M/L$ ratios lie below the universal
value because galaxies must be overrepresented in dense regions to
match the observed clustering.  Conversely, cluster $M/L$ ratios lie
above the universal value for higher $\sigma_8$.

Averaging our results over the same masses as the CNOC cluster sample,
our results are well described by the
relations

\begin{eqnarray}
(M/L_{18})_{\rm cl} &=& 577 \left({\sigma_8 \over 0.9}\right)^{1.7}
                            \left({\Omega_m \over 0.3}\right) \hbox{\hml},
			    \label{eqn:ml18} \\
(M/L_{20})_{\rm cl} &=& 907 \left({\sigma_8 \over 0.9}\right)^{2.1}
                            \left({\Omega_m \over 0.3}\right) \hbox{\hml},
			    \label{eqn:ml20} 
\end{eqnarray}

\noindent for luminosity thresholds of $M_r=-18$ and $M_r=-20$,
respectively.  As discussed in \S 2, we estimate that uncertainties in
the HOD fits introduce a $\sim 10\%$ systematic uncertainty in the
normalization of these relations, though our present investigation is
not exhaustive.  The luminosity function of SDSS galaxies implies
universal mass-to-light ratios of $\langle M/L_{18}\rangle =
509$\whml\ and $\langle M/L_{20}\rangle = 923$\whml, where
$\omega_{0.3}\equiv \Omega_m/0.3$. The statistical uncertainty in
Blanton et al.'s (2003) luminosity density estimate implies a 2\%
uncertainty in $\langle M/L \rangle$.

If we combine equation~(\ref{eqn:ml18}) with the mean mass-to-light
ratio of CNOC clusters, \mlcla$ = 359 \pm 32$\hml, we obtain the
constraint

\begin{equation}
\left({\sigma_8 \over 0.9}\right) \left({\Omega_m \over 0.3}\right)^{0.6}
= 0.75 \pm 0.06,
\label{eqn:omsig8}
\end{equation}

\noindent or $\sigma_8 \Omega_m^{0.6} = 0.33 \pm 0.03$.  Here we have
added in quadrature the 9\% statistical error on the mean cluster
$M/L$, the 2\% statistical error in the mean luminosity density, and
our estimated 10\% systematic error in the normalization of
equation~(\ref{eqn:ml18}), but we have not considered other possible
sources of systematic error.  We find similar results comparing to the
$M/L$ ratios of SDSS clusters inferred from \cite{bahcall03a} or to
the $M/L$ ratios of the virial regions of clusters found by
\cite{rines04}, but the systematic uncertainties are larger and harder
to quantify because of our reliance on mean scaling relations in the
former case and the complications of passband and geometry conversions
in the latter.  If we used the \cite{rines04} $M/L$ estimates from
cluster infall regions, we would infer a somewhat lower value of
$\sigma_8\Omega_m^{0.6}$.

Our estimate (\ref{eqn:omsig8}) derived from cluster mass-to-light
ratios conflicts with recent estimates obtained by combining CMB
anisotropy measurements with the large scale galaxy power spectrum,
Type Ia supernova data, the Ly$\alpha$ forest flux power spectrum, and
weak gravitational lensing, which tend to favor $\sigma_8 \approx
0.9$, $\Omega_m \approx 0.3$ (e.g.,
\citealt{spergel03,tegmark04,seljak04}).  For example, \cite{seljak04}
quote $\sigma_8=0.897^{+0.033}_{-0.031}$,
$\Omega_m=0.281^{+0.023}_{-0.021}$ as their combined constraint for a
6-parameter, spatially flat $\Lambda$CDM model, corresponding to
$\sigma_8\Omega_m^{0.6}=0.419\pm 0.026$.  Our conclusion agrees well
with that of \cite{vdb03}, who find that ``concordance'' values of
$\sigma_8=0.9$, $\Omega_m=0.3$ are favored by their conditional
luminosity function analyses of the 2dFGRS only for cluster
mass-to-light ratios of $750$\hml.  If they impose a more
observationally plausible constraint of \mlcl$=350 \pm 70$\hml\ (in
$b_J$-band), they find $\Omega_m=0.25^{+0.10}_{-0.07}$ and
$\sigma_8=0.78\pm 0.12$, in good agreement with
equation~(\ref{eqn:omsig8}).  Although our calculation is similar in
overall concept to van den Bosch et al.'s, we parameterize the problem
in a completely different way, fit different constraints in a
different order, use different approximations, and analyze
measurements from an independent galaxy redshift survey, red-selected
instead of blue-selected. In contrast to our fixed linear $P(k)$,
\cite{vdb03} vary the power spectrum shape parameter linearly with
$\om$. The agreement of the two results is therefore a good indication
that the conclusions are robust to details of the measurements or
analysis procedures.

Equation~(\ref{eqn:omsig8}) is nearly identical to the constraint
$\sigma_8\Omega_m^{0.6}=0.33 \pm 0.03$ obtained by \cite{bahcall03b}
in their analysis of the mass function of clusters in the SDSS Early
Data Release.  Earlier analyses of cluster mass functions have
generally yielded higher normalizations of this constraint (e.g.,
\citealt{white93,eke96,bahcall98,henry00}).  While the mass function and
mass-to-light ratio methods both incorporate cluster masses, they are
physically distinct: the former has no dependence on galaxy
luminosities, and the latter uses an average mass and is therefore
insensitive to scatter in observational estimates. Low values of
$\sigma_8$ or $\Omega_m$ would also help explain observational estimates
of the galaxy pairwise velocity distribution, which appear to conflict
with predicted values for $\sigma_8=0.9$, $\Omega_m=0.3$ \citep{Yang04}.

Our HOD modeling assumes that satellite galaxies in halos have a radial
profile corresponding to an NFW model with the concentration predicted
for $\Omega_m=0.3$.  To test our sensitivity to this assumption, we
lower all of the assumed halo concentrations in the $\sigma_8=0.9$ model
by 30\% and refit \wp, which results in slightly different $P(N|M)$ and
thus slightly different $M/L$.  We find a $1.8\%$ difference in \mlcl\
for this HOD model, so radial profiles are not an important source of
systematic uncertainty.  As noted in \S 2, changing the assumed linear
matter power spectrum from the \cite{ebw} parameterization to a CMBFAST
calculation makes a significant difference to the $\chi^2$ values of
\wp\ fits but minimal change to the HOD parameters themselves, so our
$M/L$ predictions are not sensitive to modest changes in the power
spectrum shape. The main source of systematic uncertainty is our \mlcl\
predictions is therefore the 10\% uncertainty associated with the HOD
parameterization and the
\wp\ measurements, discussed at the end of \S 2 (see
Fig. \ref{vary_alpha}).

Turning to the observational uncertainties, cluster $M/L$ ratios could
be underestimated if masses are biased low or luminosities are biased
high.  Cluster mass estimation is a challenging problem, but the
generally good agreement between virial mass estimates, Jeans equation
estimates, and estimates from X-ray or weak lensing data (see Bahcall
\& Comerford [2002] and references therein) argues against a
systematic error as large as a factor of $588/359=1.64$, the ratio of
our predicted $M/L$ for $\Omega_m=0.3$, $\sigma_8=0.9$ to Carlberg et
al.'s (\citeyear{carlberg96}) observational estimate.  Luminosities
could be biased high if background subtraction methods do not
adequately account for overdense structures surrounding clusters, an
issue that warrants further investigation with realistic mock galaxy
catalogs.  We see no obvious holes in the CMB + large scale structure
analyses, but in contrast to the approach taken here and in
\cite{vdb03}, these methods of inferring cosmological parameters rely
on a detailed theoretical model of primordial fluctuations and their
linear evolution, and they are sensitive to quantities like the
electron scattering optical depth, the CMB tensor-to-scalar ratio, and
the curvature of the inflationary fluctuation spectrum.

At present, it seems plausible that the results of these different
methods can be reconciled without major revisions, if $\sigma_8$ and
$\Omega_m$ lie at the low end of the ranges allowed by
\cite{tegmark04} or \cite{seljak04}, cluster $M/L$ ratios are somewhat
higher than the estimates shown in Figures~\ref{ml_halo} and~\ref{ml_cluster},
and our HOD parameterization leads to a modest over-prediction of $M/L$
for a given $\sigma_8$ and $\Omega_m$.  Fortunately, the remaining
uncertainties can be substantially reduced in the near future.
The SDSS redshift survey is now large enough that rich clusters
can be identified directly from the redshift survey itself,
reducing (though not eliminating) problems of contamination and
background subtraction.  Follow-up observations of these systems
can provide consistency checks of mass estimates via galaxy dynamics,
X-ray modeling, and weak lensing.  The HOD parameter constraints can
be greatly improved by bringing in additional clustering measurements,
most notably the group multiplicity function (A.\ Berlind et al.,
in preparation) and the projected three-point correlation function.
Redshift-space distortion analysis can yield independent constraints
on $\sigma_8$ and $\Omega_m$.  Advances on all of these fronts should
soon show whether the current tension in parameter estimates arises
from an accumulation of systematic errors or instead signals the need
for a new physical ingredient in the standard cosmological scenario.

We thank Frank van den Bosch, Risa Wechsler, and Jaiyul Yoo for helpful
discussions.  This work was supported by NSF Grant AST-0407125. JT
acknowledges the support of a Distinguished University Fellowship at
Ohio State University, and ZZ acknowledges the support of NASA through
Hubble Fellowship grant HF-01181.01-A awarded by the Space Telescope
Science Institute, which is operated by the Association of Universities
for Research in Astronomy, Inc., for NASA, under contract NAS
5-26555. The simulations were performed on the Beowulf and Itanium
clusters at the Ohio Supercomputing Center under grants PAS0825 and
PAS0023.  Funding for the creation and distribution of the SDSS Archive
has been provided by the Alfred P. Sloan Foundation, the Participating
Institutions, the National Aeronautics and Space Administration, the
National Science Foundation, the U.S. Department of Energy, the Japanese
Monbukagakusho, and the Max Planck Society.  The Participating
Institutions are listed at the SDSS web site, {\tt http://www.sdss.org}.


\appendix

\section{Halo and Galaxy Bias}

Halo clustering is biased relative to that of the underlying mass
distribution, by an amount that depends on halo mass. Halo bias
factors are an important ingredient in analytic calculations of galaxy
clustering, including those in this paper. Following the pioneering
work of \cite{mowhite96}, a number of authors have investigated halo
bias using N-body simulations (e.g., \citealt{porciani99,
shethlemson99, Sheth99, jing98, jing99, smt01}). However, many of
these studies are based on simulations of either low mass resolution
or limited box size (the main exception is the recent study of Seljak
\& Warren [2004], which we discuss below). Most previous studies have
also compared different cosmologies such as open and standard CDM to
one version of \lcdm, rather than focusing on variants of the \lcdm\
cosmology with a wide range of $\s8$ as we have done here. We have
also performed an identical set of simulations with a power spectrum
shape parameter of $\Gamma=0.12$, which significantly increases the
large-scale power and reduces the power on small scales relative to
our standard choice of $\Gamma=0.2$. Our simulations are therefore
well suited to investigate halo bias for the cosmological models of
the greatest interest today, and to investigate the dependence of
bias factors on power spectrum shape or normalization. The use of five
realizations and a reasonably large volume (253 \hmpc\ on a side)
yields good statistics for high mass halos. Note that we define halos
using a friends-of-friends algorithm with linking length
$l=0.2n^{-1/3}$. Alternative definitions would yield slightly
different halo masses (see \citealt{hukravtsov03}) and would therefore
require a slightly different formula for $b_h(M)$.

The halo bias factor can be defined by the ratio of halo and mass
autocorrelations, or power spectra, or using the halo--mass
cross-correlation. Since we are interested in modeling galaxy
autocorrelations, we adopt the definition $b^2_h(M) =
\xi_h(r,M)/\xi_m(r)$, where $\xi_h(r,M)$ is the autocorrelation
function of halos of mass $M$ and $\xi_m(r)$ is the non-linear matter
correlation function measured from the simulations. Sheth \etal\
(2001, hereafter SMT) give an analytic formula for $b_h$, motivated by
the analytic model of \cite{Sheth99} but empirically calibrated
against numerical simulations:

\begin{equation}
\label{e.smt}
b_h(\nu) = 1 + \frac{1}{\sqrt{a}\delta_c} \left[ \sqrt{a}(a\nu^2) + \sqrt{a}b(a\nu^2)^{1-c} 
  - \frac{(a\nu^2)^c}{(a\nu^2)^c + b(1-c)(1-c/2)} \right],
\end{equation}

\noindent where $\delta_c=1.686$ is the critical overdensity required
for collapse and $\nu=\delta_c/\sigma(M)$, with $\sigma(M)$ the
linear theory rms mass fluctuation in spheres of radius
$r=(3M/4\pi\bar{\rho})^{1/3}$. The three parameters in this equation
are $a=0.707$, $b=0.5$, and $c=0.6$, as listed in SMT.

\begin{figure}
\epsscale{1.0}
\plotone{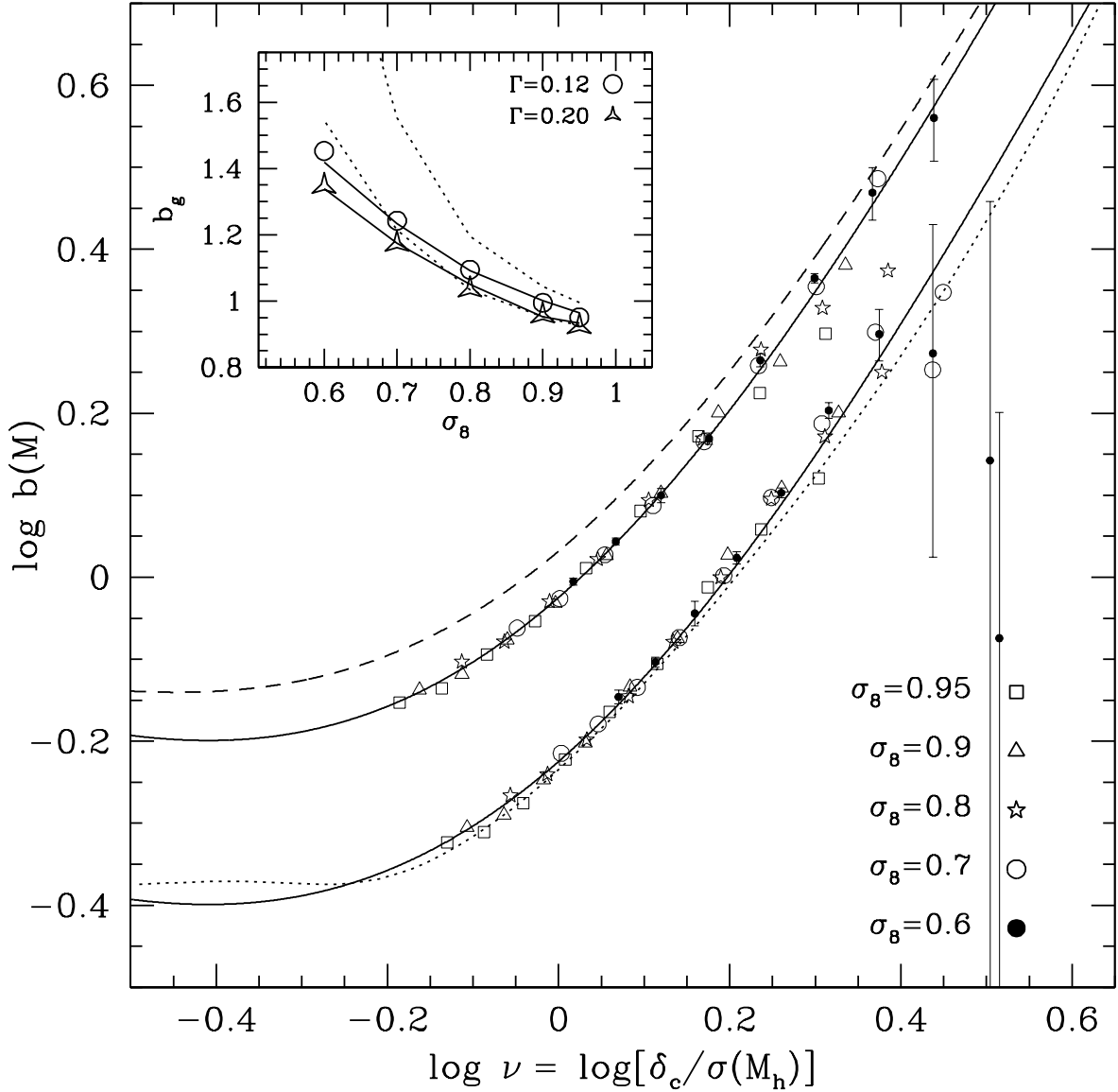}
\caption{ \label{halo_bias} The halo bias factor is shown as a
  function of $\nu$ for the five different values of $\s8$ for both
  $\Gamma=0.2$ and $\Gamma=0.12$. The values of the bias factor were
  calculated from the asymptotic value of the correlation function at
  large scales: $b_h = \sqrt{\xi_h/\xi_m}$. The $\Gamma=0.12$ results
  are offset by 0.2 dex to avoid crowding. The dashed line is the bias
  relation given by \cite{smt01}. The solid line shows the bias
  relation of the same functional form, but using the parameters
  $a=0.707$, $b=0.35$, and $c=0.80$. The dotted line is the bias
  relation of \cite{seljakwarren04} calculated for $\s8=0.9$ and
  $\Gamma=0.2$.  The error bars, shown only for $\s8=0.6$, are the
  error in the mean of the five realizations.  {\it Inset box:} The
  galaxy bias parameter for the HOD models of \cite{tinker04}, plotted
  as a function of $\s8$. The solid lines are analytic calculations of
  the galaxy bias using the new bias relation. The dotted lines are
  the analytic calculations of $b_g$ using the bias relation of
  \cite{seljakwarren04}.}
\end{figure}

We divide the halos in our simulations into bins separated by factors
of two in mass. The logarithmic center of the lowest bin is
$1.22\times 10^{12} (\om/0.3)$ \hmsol, below $M_\ast$ for $\s8>0.$6.
We calculate halo bias by averaging $\sqrt{\xi_h/\xi_m}$ for radii
$4\leq r \leq 12$ \hmpc, a regime in which the ratio of the halo and
matter correlation functions is approximately constant and noise is
not a factor. The results are plotted in Figure \ref{halo_bias}. The
dashed line, which shows the SMT bias relation, is significantly
higher than the values of $b_h$ calculated from the simulations for
both values of $\Gamma$. A better fit to the calculations is shown
with the solid line, which plots a bias relation of the same form as
equation (\ref{e.smt}), but with $a=0.707$, $b=0.35$, and
$c=0.80$. With these parameter values, the formula gives accurate fits
to our numerical results for the full range of $\sigma_8$ values, and
it works equally well for $\Gamma=0.2$ and $\Gamma=0.12$. Increasing
either the lower or upper bounds of the radial range over which $b_h$ is
calculated does not appreciably change the bias values or the quality
of the fits.

The inset box in Figure \ref{halo_bias} plots an example of using
equation (\ref{e.bg}) to calculate the galaxy bias with our modified
parameters of equation (\ref{e.smt}), which we compare to the galaxy
bias in the simulations calculated by the same method as the halo bias
for five values of $\s8$. The plot symbols represent the galaxy bias
calculated from the simulations, and the solid lines represent the
analytic calculations (eq. [\ref{e.bg}]) using equation (\ref{e.smt})
with our new parameters. For this test, the HOD parameters used to
populate the simulations are taken from Table 1 of \cite{tinker04},
parameters similar to the $M_r<-20$ sample. The analytically calculated
bias factors differ from the simulation results by $\lesssim 1$\%.

Recently Seljak \& Warren (2004, hereafter SW) proposed a new
empirically determined halo bias relation, empirically calibrated on
large simulations (up to $768^3$ particles) with cosmological parameters
close to the best-fit values from CMB and large-scale structure
measurements. The dotted line in Figure \ref{halo_bias} shows the SW
formula evaluated for $\Gamma=0.2$ and $\s8=0.9$, which fits our
numerical data accurately. However, the SW formula is expressed in terms
of $M/M_\ast$, where $\sigma(M_\ast)=\delta_c$, instead of
$\nu=\delta_c/\sigma(M)$, and the mapping between $M/M_\ast$ and $\nu$
depends on the amplitude and shape of the power spectrum. We find that
the SW formula becomes a poor fit to our results at low $\sigma_8$, and
the discrepancies are worse for $\Gamma=0.12$, as illustrated in the
inset box by calculating the galaxy bias parameters with the SW halo
bias formula. Thus, their formula is accurate for models close to the
cosmological concordance model, but our modified version of the SMT
formula, with $\nu$ as the halo mass parameter, applies more
universally.

We have not investigated alternative definitions of bias using the
power spectrum or mass cross-correlation, or using linear instead of
non-linear matter clustering. We also note that our numerical results
do not extend much below $\nu=1$, and simulations of smaller volumes
or larger dynamic range are needed to test equation (\ref{e.smt}) in
the low mass regime.


\section{The Analytic Model}

Our analytic calculation of the two-point galaxy correlation function
is similar to that presented by \cite{zheng04a} and used in Zehavi
\etal's (2004a,b) modeling of the SDSS \wp. Here we report
improvements in the procedure that have been incorporated in our
present analysis. For completeness and clarity, we describe the method
from start to finish. The new ingredients are the use of the modified
halo bias formula of Appendix A and the more accurate treatment of
halo exclusion described by equations (\ref{e.pgg_sph}) and
(\ref{e.pgg_ell}). The \cite{zheng04a} procedure was calibrated and
tested using the 144 \hmpc, $256^3$ particle GIF simulation of
\cite{jenkins98}, which was also used to calibrate the SMT
formula. Our use of the five 253 \hmpc, $360^3$ particle simulations
described in \S 3 allows us to achieve a more accurate calibration and
to test the procedure for a range of cosmological parameters.

The correlation function is defined as the probability above random of
there being a pair of objects at separation $r$. In the HOD context,
a pair of galaxies can reside within a single halo or come from two distinct
halos. These two contributions are computed separately and combined
to get the full correlation function, i.e.

\begin{equation}
\xi(r) = [1+\xi_{\rm 1h}(r)] + \xi_{\rm 2h}(r).
\end{equation}

\noindent The ``1+'' arises because it is the pair counts, proportional to
$1+\xis$ and $1+\xid$, that sum to give the total pair counts,
proportional to $1+\xi$. The one-halo term is calculated in real space
through (Berlind \& Weinberg 2002)

\begin{equation}
1+\xis(r)=\frac{1}{2\pi r^2\ngavg^2}
              \intdn\frac{\langle N(N-1)\rangle_M}{2}
              \frac{1}{2\Rvir(M)} F^\prime\left(\frac{r}{2\Rvir}\right),
\end{equation}

\noindent where $\ngavg$ is the mean number density of galaxies,
$dn/dM$ is the halo mass function (\citealt{Sheth99,Jenkins01}), and
$\langle N(N-1)\rangle_M/2$ is the average number of pairs in a halo
of mass $M$. The function $F(x)$ is the average fraction of galaxy
pairs in a halo of mass $M$ (or virial radius $\Rvir$) that have
separation less than $r$, which is related to the halo density
profile, $\rho_m(r)$, and $F^\prime(x)$ is its derivative. In practice,
$F(x)$ must be treated differently for central-satellite galaxy pairs
and satellite-satellite pairs. In the former, the pair distribution is
proportional to the volume-weighted density profile, $F^\prime(x)
\propto \rho_m(r)r^2$, normalized to one. For the latter it is derived
from the halo profile convolved with itself, a calculation that can be
done analytically for an NFW profile (\citealt{Sheth01a}). The average
number of one-halo pairs in the range $(x, x+dx)$ in halos of mass
$M$ can be written explicitly as

\begin{equation}
        \frac{\langle N(N-1)\rangle_M}{2}F^\prime(x)\,dx = \langle N_{\rm
        cen}N_{\rm sat} \rangle_MF^\prime_{\rm cs}(x)\,dx + \frac{\langle
        N_{\rm sat}(N_{\rm sat}-1)\rangle_M}{2}F^\prime_{\rm ss}(x)\,dx,
\end{equation}

\noindent where the subscripts $cs$ and $ss$ refer to
central-satellite pairs and satellite-satellite pairs
respectively. For a Poisson distribution of satellite occupation,
$\langle N_{\rm sat}(N_{\rm sat}-1)\rangle=\langle N_{\rm
sat}\rangle^2$. We use an NFW profile with concentration parameters as
a function of halo mass calculated by the method of \cite{bullock01}
and \cite{kuhlen04}.

The one-halo term dominates \xg\ at small scales, while the two-halo
term fully accounts for all galaxy pairs at separations $\gtrsim 5$
\hmpc. The transition region between one-halo and two-halo dominance
is difficult to model because only certain regions of the halo mass
function can contribute to the two-halo term at small scales. The
range of halo masses included must ensure that halo pairs do not
overlap, since such halo pairs would be merged into a single halo by
the friends-of-friends scheme that we use to define halos in the first
place. It is this one-halo to two-halo transition that causes \wp\ to
deviate from a power-law at scales near 1 \hmpc\
(\citealt{zehavi04a,zehavi04b}); as $r$ increases, $\xis$ drops
rapidly, while the rise in $\xid$ is regulated by halo exclusion. For
brighter galaxies, which preferentially occupy high-mass halos, the
rise in $\xid$ occurs at a larger $r$, making the deviation from a
power-law greater for brighter galaxy samples.

Since the radial distribution of galaxies within halos must be
accounted for in the calculation of the two-halo term, the calculation
itself is done in Fourier space, where the convolutions with the halo
density profile become multiplications instead
(\citealt{scherrer91,seljak00,roman01}). In the method of
\cite{zheng04a}, halo exclusion was treated by only including halos
with virial radii less than half the value of $r$ for which \xg\ is
being calculated, i.e.\ with masses below

\begin{equation}
\label{e.mlim}
M_{\rm lim} = \frac{4}{3}\,\pi\, \left(\frac{r}{2}\right)^3\, \pcrit\, \om\, \Delta,
\end{equation}

\noindent where $\Delta$ is the virial overdensity of the halo,
relative to the mean density, which we have chosen to be 200. With
this implementation of halo exclusion, the calculation of the two-halo
term in Fourier space is

\begin{equation}
\label{e.pgg}
P_{\rm gg}^{\rm 2h}(k,r)=P_m(k)\left[\frac{1}{\ngavgp}\intdnM 
                       \navg b_h(M,r) y_g(k,M)\right]^2, 
\end{equation}

\noindent where $y_g(k,M)$ is the Fourier transform of the halo
density profile (e.g. \citealt{cooraysheth02}), $b_h(M,r)$ is the halo
bias at separation $r$, and the restricted number density $\ngavgp$ is
the average number density of galaxies that reside in halos with
$M\le\mlim$,

\begin{equation}
\label{e.ngavgp}
\ngavgp = \intdnM \navg.
\end{equation}

\noindent The matter power spectrum, $P_m(k)$, is the non-linear form
given by \cite{smith03}. At large $r$, equation (\ref{e.pgg}) can be
thought of as simply multiplying the non-linear matter power spectrum
by the galaxy pair-weighted halo bias factor to obtain the galaxy
power spectrum. At smaller separations the finite size of the halos
must be taken into account. The scale dependence of halo bias also
becomes important at $r \lesssim 3$ \hmpc. Parameterizing the scale
dependence by the amplitude of the non-linear matter correlation
function, the scale dependence of halo bias is well described by

\begin{equation}
\label{e.scale_bias}
  b^2(M,r)=b^2(M)\, \frac{[1+1.17\,\xi_m(r)]^{1.49}}{[1+0.69\,\xi_m(r)]^{2.09}},
\end{equation}

\noindent where $b(M)$ is the large-scale bias, for which we have used
the bias relation given in Appendix A. Equation (\ref{e.scale_bias}),
determined from inspection of our numerical simulations, is fairly
accurate for the full range of $\s8$ values and both values of
$\Gamma$ explored in this paper.

At a given $r$, equation (\ref{e.pgg}) is solved for all $k$, then
converted to real space by

\begin{equation}
\label{e.pk2xi}
\xid^\prime(r)=\frac{1}{2\pi^2}\int_0^\infty P_{\rm gg}^{\rm 2h}(k,r) k^2 
        \frac{\sin kr}{kr} dk,
\end{equation}

\noindent where $\xid^\prime(r)$ denotes that we have calculated the
two-halo term for a restricted range of the halo mass function.
The value calculated in equation (\ref{e.pk2xi}) is converted to a
probability over random for the entire halo (and therefore galaxy)
population by

\begin{equation}
\label{e.xi2h_correction}
1+\xid(r) = \left( \frac{\ngavgp}{\ngavg}\right)^2 [1+\xid^\prime(r)].
\end{equation}

\begin{figure}
\epsscale{1.0}
\plotone{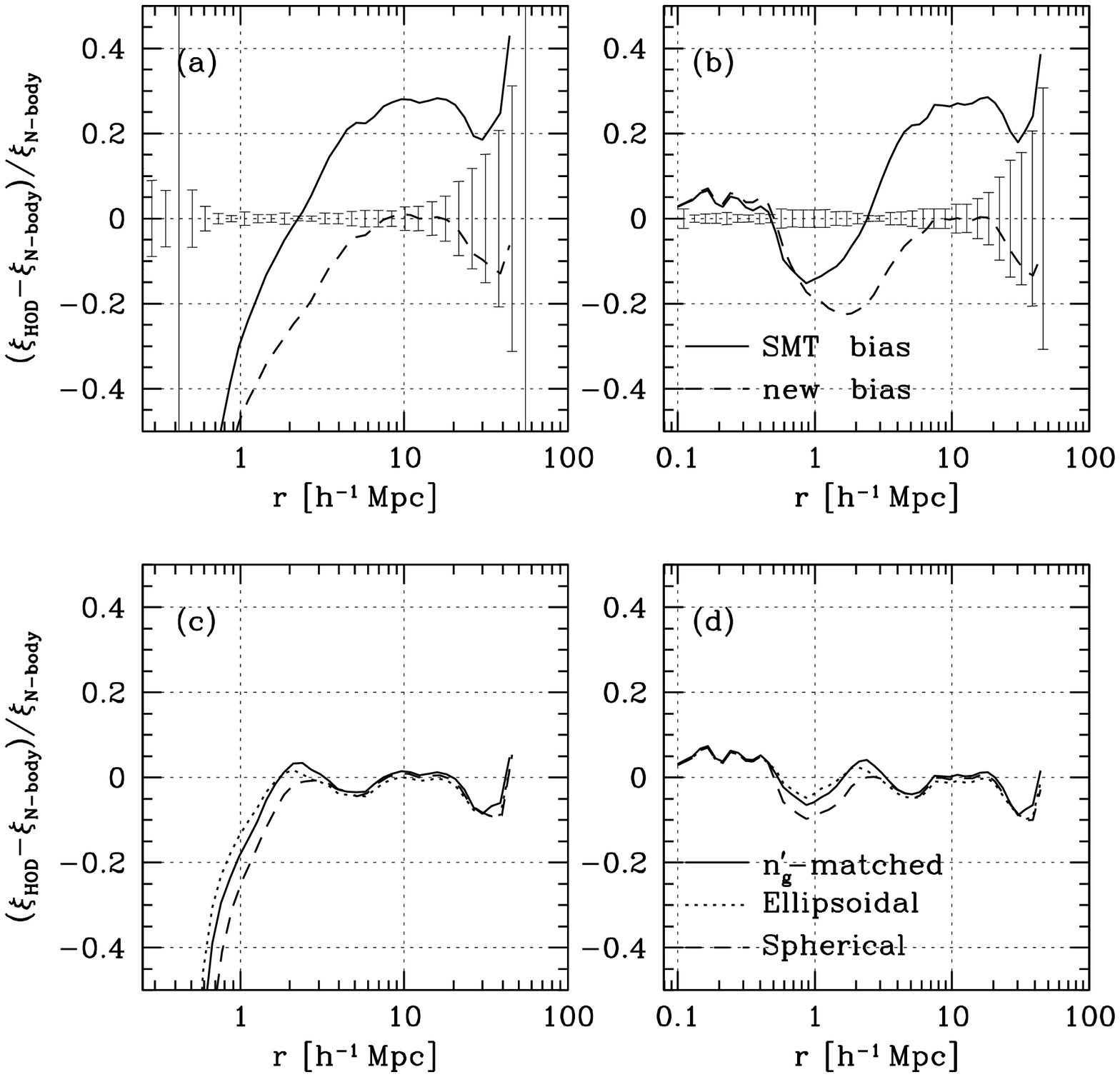}
\caption{ \label{hod_model} The analytic HOD calculation of \xg\ is
  compared to numerical results for different bias and halo exclusion
  prescriptions. Panel (a): The calculation of $\xid$ using the halo
  exclusion approach of \cite{zheng04a} is compared to N-body
  results. The solid line the calculation performed with the halo bias
  prescription of \cite{smt01}. The dashed line is the same
  calculation using the modified bias in Appendix A. The error bars
  represent the error in the mean from the five realizations. Panel
  (b) compares the total calculation of \xg\ with the numerical
  results. Panels (c) and (d) compare $\xid$ and \xg\ calculated by
  the three new halo exclusion methods described in Appendix B to the
  N-body results. }
\end{figure}

The virtue of this implementation is that the integral over mass in
equation (\ref{e.pgg}) is calculated once and squared, instead of
being a double integral over different halo pairs. This approximation,
however, neglects galaxy pairs from halos larger than $\mlim$ paired
with smaller halos. In Figure \ref{hod_model}, we compare this
analytic method to numerical results. For this comparison we use the
correlation functions calculated from the N-body simulations described
in \S 3, but following Tinker et al.'s (2004) practice of drawing
satellite galaxy populations from the appropriate NFW profile instead
of randomly sampling the friends-of-friends dark matter distribution,
as done in \S 3.  We use the $\s8=0.8$ output with HOD parameters
$\mmin=1.11\times 10^{12}$, $M_1=2.53\times 10^{13}$, and
$\alpha=1.01$.

Figure \ref{hod_model}a compares the N-body two-halo term to equation
(\ref{e.xi2h_correction}) calculated both with our new halo bias
parameters and with the original parameters of the SMT function. At
$r\gtrsim 10$ \hmpc, the original SMT relation over-predicts the
galaxy bias by $\sim 15\%$ and the correlation function by $\sim
30\%$. The new bias function yields an excellent match at these
scales. At smaller scales, this method of halo exclusion
under-predicts the number of two-halo pairs, with errors greater than
50\% at 1 \hmpc. Although the one-halo term begins to dominate at this
scale, the large error in $\xid$ is still apparent in the total
correlation function, shown in panel (b). The $\sim 5\%$ error at the
smallest scales is due to the fact that the halo mass function of our
simulations is not precisely represented by the \cite{Jenkins01}
function assumed in the analytic calculation. The two-halo term is
much less sensitive to the mass function, and is not affected by
this difference.

The numerical test in the Appendix of \cite{zehavi04a} showed a much
smaller discrepancy on large scales because the 144 \hmpc\ GIF
simulation used for the test (and for the calibration of the SMT bias
factors) has a high amplitude of large scale clustering for low mass
halos. With our multiple, larger volume simulations, the need for
lower halo bias factors is evident, and these in turn drive the need
for a more accurate treatment of halo exclusion. Zehavi et
al. (2004a,b) find a low $\chi^2/{\rm d.o.f.}$ fitting the projected
correlation function of luminous, $M_r\le -21$ galaxies, while we find
a relatively high $\chi^2/{\rm d.o.f.}$ for this sample (see Table
1). However, as noted in \S 2, the $\chi^2$ values of \wp\ fits are
sensitive to the difference between a $\Gamma=0.2$ power spectrum and
a (presumably more realistic) CMBFAST power spectrum, even though
the best-fit HOD parameters are not. Combination of our present \xg\
calculation with the CMBFAST power spectrum produces a similar \wp\
for $M_r\le-21$ galaxies, with similarly low $\chi^2$, to that
obtained with the \cite{zheng04a} prescription and a $\Gamma=0.2$
power spectrum. Thus, this more accurate modeling leads to the same
bottom line conclusion as Zehavi et al. (2004a,b), and similar HOD
parameters.

We now return to the halo exclusion problem. Under the assumption of
spherical halos, all two-halo pairs would be accounted for by summing
all the galaxies from halo pairs for which the sum of virial radii is
smaller than the separation, i.e. $R_{\rm vir1} + R_{\rm vir2} \le
r$. For this ``spherical halo'' exclusion, equation (\ref{e.pgg}) must
be modified to

\begin{eqnarray}
\label{e.pgg_sph}
P_{\rm gg}^{\rm 2h}(k,r)& = & P_m(k)\,\frac{1}{\ngavgp{^2}}\int_{0}^{\mlima} dM_1\frac{dn}{dM_1}
                       \navga b_h(M_1,r) y_g(k,M_1) \nonumber \\
		      & &  \int_{0}^{\mlimb} dM_2\frac{dn}{dM_2}
                       \navgb b_h(M_2,r) y_g(k,M_2),
\end{eqnarray}

\noindent where $\mlima$ is the maximum halo mass such that
$\Rvir(\mlima)=r-\Rvir(\mmin)$ and $\mlimb$ is related to $M_1$ by
$\Rvir(\mlimb)=r-\Rvir(M_1)$. Since the upper limit of the second
integral depends on the integrand of the first, equation
(\ref{e.pgg_sph}) must be solved as a double integral, making it more
computationally expensive than equation (\ref{e.pgg}), but
significantly more accurate by increasing the number of
small-separation two-halo pairs counted.

The range of halo masses over which the two-halo term is calculated is
different than that in equation (\ref{e.pgg}), and the restricted
number density $\ngavgp$ must reflect that change. Using this new
method of halo exclusion, $\ngavgp$ becomes

\begin{equation}
\label{e.ng_sph}
\ngavgp{^2} =  \int_{0}^{\mlima} dM_1\frac{dn}{dM_1}
   \navga \,\int_{0}^{\mlimb} dM_2\frac{dn}{dM_2} \navgb.
\end{equation}

\noindent Figures \ref{hod_model}c and \ref{hod_model}d compare our
numerical results to the spherical exclusion method. At $r=1$ \hmpc,
where the previous method resulted in a $\sim 50\%$ error, the error
has been reduced to $\sim 25\%$. At $r=2$ \hmpc, the $\sim 20\%$ error
in the previous method is eliminated entirely.

Halos are not spherical objects, however. They are triaxial objects
that can exhibit significant flattening (see,
e.g. \citealt{jingsuto02}). This can lead to halo pairs which are
closer than the sum of their virial radii. By assuming a lognormal
distribution of ellipticities with mean flattenings motivated by
simulation results, one can determine the probability that halos of a
given separation are allowed. This probability, $P(x)$, where
$x=r/(R_{\rm vir1}+R_{\rm vir2})$, can be used to modify the two-halo
calculation to 

\begin{eqnarray}
\label{e.pgg_ell}
P_{\rm gg}^{\rm 2h}(k,r)& = & P_m(k)\,\frac{1}{\ngavgp{^2}}\int_{0}^{\infty} dM_1\frac{dn}{dM_1}
                       \navga b_h(M_1,r) y_g(k,M_1) \nonumber \\
		      & &  \,\int_{0}^{\infty} dM_2\frac{dn}{dM_2}
                       \navgb b_h(M_2,r) y_g(k,M_2) P(x),
\end{eqnarray}

\noindent where the limits in equation (\ref{e.pgg_ell}) are both
infinity (but in practice can be cut off at some reasonably large
value) but the calculation is still a double integral because the value
of $M_1$ is used in the ellipsoidal exclusion probability in the
integral over $M_2$.

We investigated the ellipsoidal exclusion probability by a Monte
Carlo approach. We assume a lognormal distribution of halo axis ratios q
with dispersion 0.2 and means $q_b = b/a=0.9$ and $q_c = c/a=0.8$,
motivated by the results of \cite{jingsuto02}, then assume the axis
ratios have a lognormal distribution in $q$ with a dispersion of 0.2.
By randomly selecting ellipticities and orientation angles for halos of
a given mass ratio and separation, we find that the probability of
non-overlapping halos is well approximated by $P(y) = (3y^2 - 2y^3)$ in
the range $0 \le y \le 1$, where $y = (x-0.8)/0.29$. At $y<0$, $P(y)=0$,
and at $y>1$, $P(y)=1$. The restricted number density, $\ngavgp$, must
also be calculated in this way. Equation (\ref{e.ngavgp}) becomes

\begin{equation}
\ngavgp{^2} =  \int_{0}^{\infty} dM_1\frac{dn}{dM_1}
   \navga \,\int_{0}^{\infty} dM_2\frac{dn}{dM_2}
   \navgb  P(x).
\end{equation}

\noindent The results of the ellipsoidal halo exclusion are presented
in panels (c) and (d) of Figure \ref{hod_model}. The ellipsoidal
exclusion approach is an improvement over spherical exclusion; at $r=1$
\hmpc, the error in the two-halo term has been reduced to $\sim
10\%$. Choosing more extreme values of $q_b$ and $q_c$ does not
significantly change the results. This ellipsoidal exclusion method is
the one we have used for the calculations in this paper.

Although this approach is more accurate, the added computation time can
become prohibitive when fitting observed data to high precision, where
many hundreds of iterations are required. We have therefore created a
halo exclusion approach that mimics the results of ellipsoidal exclusion
but can be written as separable integrals in the Fourier-space
calculation of $P_{\rm gg}^{\rm 2h}(k,r)$. Although the evaluation of
equation (\ref{e.pgg_ell}) is CPU intensive, the calculation of the
restricted number density under the ellipsoidal approach is relatively
rapid. For a more efficient scheme, we recalculate the ellipsoidal
$\ngavgp$ first, and then $\mlim$ in equation (\ref{e.ngavgp}) is
increased until the restricted number density matches that of the
ellipsoidal calculation. We then use equation (\ref{e.pgg}) to calculate
$\xid$.

The results of this approximation, which we call $\ngavgp$-matched,
are shown in panels (c) and (d) of Figure \ref{hod_model}. The loss of
accuracy relative to the full ellipsoidal treatment is minimal. In
further tests with multiple values of $\s8$ and HOD parameters that
give higher and lower galaxy space densities, we find similar results.

Although the methods introduced here significantly improve the
analytic calculation of \xg\ for specified cosmological and HOD
parameters, there is room for further investigation and
improvement. Outstanding issues include the small but non-negligible
dependence of the halo mass function shape on cosmology (i.e., the
non-universality of the Jenkins et al. [2004] formula), the effect of
scatter in concentrations and halo ellipticity on the one-halo term,
bias factors of low mass halos, effects of cosmology on the
scale-dependence of halo bias, the dependence of the approximation on
the halo definition, and the interaction of all of these effects with
the treatment of halo exclusion. We are presently investigating a
number of these issues. The long-term goal is to ensure that errors in
the calculation of \xg\ for specified parameters are a negligible
source of uncertainty in the inference of HOD and cosmological
parameters from observational data. Because of the high precision of
the clustering measurements, it is not clear that we have yet reached
this goal.


{}


\end{document}